\documentclass[a4paper,fleqn,usenatbib]{mnras}

\usepackage{newtxtext,newtxmath}

\usepackage[T1]{fontenc}
\usepackage{ae,aecompl}

\usepackage{graphicx}	
\usepackage{amsmath}	
\usepackage{amssymb}	

\newcommand{\kms}{\,km\,s$^{-1}$}
\newcommand{\kmsperkpc}{\,km\,s$^{-1}$\,kpc$^{-1}$}
\newcommand{\msun}{\,M$_\odot$}

\newcommand{\dkin}{\,$D_{\mathrm{kin}}$}
\newcommand{\dact}{\,$D_{\mathrm{actual}}$}

\title[Streaming Motions and Kinematic Distances]{Streaming Motions and Kinematic Distances to Molecular Clouds}

\author[F. G. Ram\'on-Fox et al.]{
F. G. Ram\'on-Fox \thanks{E-mail: fgr2@st-andrews.ac.uk}
and Ian A. Bonnell
\\
Scottish Universities Physics Alliance (SUPA), School of Physics and Astronomy, University of St. Andrews, \\ North Haugh, St. Andrews, Fife KY16 9SS, UK
}

\date{Accepted XXX. Received YYY; in original form ZZZ}

\pubyear{2017}

\begin{document}
\label{firstpage}
\pagerange{\pageref{firstpage}--\pageref{lastpage}}
\maketitle

\begin{abstract}
We present high-resolution smoothed particle hydrodynamics simulations of a region of gas flowing in a spiral arm and identify dense gas clouds to investigate their kinematics with respect to a Milky Way model. We find that, on average, the gas in the arms can have a net radial streaming motion of $v_R \approx -9$ \kms~and rotate $\approx 6$\kms~slower than the circular velocity. This translates to average peculiar motions towards the Galaxy centre and opposite to Galactic rotation. These results may be sensitive to the assumed spiral arm perturbation, which is $\approx 3\%$ of the disc potential in our model. We compare the actual distance and the kinematic estimate and we find that streaming motions introduce systematic offsets of $\approx 1$ kpc. We find that the distance error can be as large as $\pm 2$ kpc and the recovered cloud positions have distributions that can extend significantly into the inter-arm regions. We conclude that this poses a difficulty in tracing spiral arm structure in molecular cloud surveys.

\end{abstract}




\section{Introduction}

Understanding star formation requires knowledge of molecular cloud properties, their formation and evolution. Observations show that molecular gas is associated with spiral arms (e.~g. \citealt{Rosolowskyetal2007,Schinnereretal2013}), which is supported by simulations as well (e.~g. \citealt{Bonnelletal2006,DobbsBonnellPringle2006}). Tracing the position of molecular gas in the Galaxy requires an accurate measurement of its distance (e.~g. \citealt{HeyerandDame2015}). 

Several methods are available for obtaining distances to molecular clouds. A widely used estimate is the kinematic distance, which assumes that clouds move on circular velocities to derive a relation between the cloud's heliocentric distance and the line of sight velocity (e.~g. \citealt{Schmidt1957, Dameetal1986, Roman-Duvaletal2009}). This method has been applied in many surveys tracing gas and star forming regions in the Milky Way (e.~g. \citealt{BlitzandSpergel1991,Kolpaketal2003,Russeiletal2011,Wienenetal2015,Raganetal2016,Riceetal2016,Miville-Deschenes-etal2017}). \citet{BruntandKerton2002} and \citet{Brunt2003} propose a distance estimation scheme based on the size--line width relation in molecular clouds. Other works use an approach based on measuring the reddening of stars on the background with respect to the cloud and compare it with foreground stars (e.~g. \citealt{Schaflyetal2014}). Trigonometric parallaxes for high-mass star forming regions offer the most reliable distances \citep{Reidetal2009}. 

A problem with tracing spiral arms in the Galaxy using kinematic distances is that the arm perturbation can introduce radial and azimuthal streaming components in the gas motions (e.~g. \citealt{Roberts1969,RobertsandStewart1987,EnglmaierandGerhard1999,Englmaier2000,Colomboetal2014}). Several works have developed different models of the velocity field of the Galaxy, which can help to incorporate the effects of streaming motions in the kinematic distance estimate \citep{BurtonandBania1974,LisztandBurton1981,FosterandMacWilliams2006,Reidetal2009,Andersonetal2012,Wienenetal2015}. 

However, when dealing with individual clouds, the local velocity field can be more complex than the global one. From observations of giant molecular clouds, \citet{StarkandBrand1989} find a cloud-to-cloud dispersion of $\approx 7.8$ \kms~for objects with distances from the Sun smaller than 3 kpc, and an average motion of $\approx 4$ \kms~with respect to the LSR. In a sample of nearby galaxies, \citet{Wilsonetal2011} find a mean dispersion of $6.1$ \kms. In simulations, \citet{Dobbsetal2015} find cloud-to-cloud dispersions in the range of $3 - 6$ \kms. Although the dispersion may be relatively small, the gas dynamics can still have net streaming motions (e.~g. \citealt{Reidetal2009}).

Simulations allow us to compare actual distances with kinematic estimates. Several works using global simulations of a Milky Way galaxy concluded that it is feasible to reconstruct the structure to some extent using kinematic distances (e.~g. \citealt{Gomez2006, Pohletal2008, Babaetal2009}). However, they find that errors can be much larger than $1$ kpc. The morphological features recovered near a spiral arm show substantial distortions with some spurious effects \citep{Gomez2006,Babaetal2009}.

In this paper, we present high resolution hydrodynamical simulations of a region of gas flowing into a spiral arm. We identify dense gas structures and obtain their velocities with respect to the galaxy. Our results show that clouds have significant streaming motions, which introduce non-negligible systematic errors in the kinematic distance. We quantify the typical error as a function of the clouds' velocity deviation and we present results in the context of dynamical simulations of the model galaxy. In \S \ref{sec:methodology}, we present a brief summary of the numerical simulations. In \S \ref{sec:cloud-kinematics}, we present results of the cloud streaming motions. In \S \ref{sec:KD-errors}, we first analyse how a cloud-to-cloud dispersion with respect to circular orbit propagates into errors in the kinematic method; then we use the simulations to show how net streaming motions introduce systematic offsets in the distance estimate. In \S \ref{sec:discussion} we discuss our results in the context of other works and \S \ref{sec:conclusions} outlines our conclusions.

\section{Methods and Simulations}
\label{sec:methodology}

For our simulations, we use the smoothed particle hydrodynamics (SPH) method \citep{GingoldMonaghan1977} to treat the gas dynamics. Our code is based on the formulation of \citet{Benz1990} and \citet{BateBonnellPrice1995}. In SPH, a gas is represented by an ensemble of particles. Each particle has a fixed mass and a smoothing length $h$, which determines the spatial resolution. The smoothing length is determined by ensuring that each particle has approximately $50$ neighbours. The code also includes an artificial viscosity for shocks with the parameters $\alpha = 1$ and $\beta = 2$ \citep{MonaghanLattanzio1985,Monaghan1992}. The thermal physics includes heating from shocks. We use the cooling function of \citet{KoyamaInutsuka2002}, which includes cooling from atomic and molecular lines as well as cooling from dust. The internal energy is integrated implicitly using the method described in \citet{Vazquez-Semadenietal2007}. For a more detailed description of the thermal physics see \citet{Bonnelletal2013} and \citet{Lucasetal2013}.

\subsection{Galaxy Model}
\label{sec:galaxy-models}

The galactic potential is represented by the combination of an axisymmetric term plus a perturbation of the spiral arms. The first component is given by a logarithmic potential (e.~g. \citealt{BandT2008}), which has a rotation curve given by:
\begin{equation}
	v_c(R) = v_0 \frac{R}{\sqrt{R_c^2 + R^2}},
	\label{eq:rot-curve}
\end{equation}
where $v_0$ is a velocity parameter, $R_c$ is a characteristic radius. This rotation curve is flat at values of $R$ larger than $R_c$. The logarithmic potential is modulated in the vertical direction by a scale factor $z_q$. These are set to $v_0 = 220$ \kms, $R_c = 0.1$ kpc, and $z_q = 0.7$ in order to produce a flat rotation curve with parameters representative of the Milky Way. 

The spiral arm perturbation uses the potential of \citet{CoxandGomez2002}, which has the functional form given by:
\begin{equation}
	\Phi(R, \phi, z) = \sum_n A_n(R,z) \cos \left(n \Gamma(r, \phi, t) \right),
\end{equation}
where $\Gamma$ is defined as:
\begin{equation}
	\Gamma(R, \phi, t) = N \left(\phi + \Omega_p t - \frac{\ln(R/R_0)}{\tan \alpha} - \phi_p \right).
\end{equation}
In these equations, $A_n(R,z)$ is associated to the strength of the perturbation, $N$ is the number of arms, $\Omega_p$ is the pattern speed, $\alpha$ is the pitch angle, $R_0$ is a characteristic radius, and $\phi_p$ is a phase offset. This produces a perturbation rotating with constant $\Omega_p$ and amplitude decreasing with distance. See \citet{CoxandGomez2002} for more details. In our simulations, the parameters are: $N = 4$, $\Omega_p = 20$ \kmsperkpc, $\alpha = 15^\circ$, $R_0 = 8$ kpc, and $\phi_p = 0$. This produces a logarithmic spiral arm with a corotation radius of $\approx 11$ kpc, given the rotation curve of the Milky Way. The amplitude is $\approx 3\%$ of the disc potential and corresponds to a change of $\approx 20$ \kms~for a particle falling in the potential. We note that we are assuming a constant spiral arm perturbation which may yield different results compared to those of transient arms obtained in $N$-body simulations (e.~g. \citealt{Babaetal2009,Pettittetal2015}). For more details, see \citet{DobbsBonnellPringle2006} and \citet{Bonnelletal2013}.

\subsection{Initial Conditions and Region of Simulation}
\label{sec:ICs}

The initial conditions for the simulations in the present work are re-sampled from the galactic scale simulation presented in \citet{Bonnelletal2013}. The region of interest is taken from a part of the simulation near a spiral arm shock. The region is sampled by $N = 31\,454\,976$ particles and the particle mass is $m_g = 0.625$\msun, thus the total mass of the region is $\approx 1.96 \times 10^7$\msun. About $53\%$ of the gas is already in the shock and spiral arm region. The remaining $47\%$ is in the inter-arm region and about to enter the spiral arm, which allows us to study the evolution of the gas falling into this region. The results in this paper are based on simulations where self-gravity is not included in order to focus on the effects of the hydrodynamics coupled to the thermal physics. This approach also allows us to follow the gas dynamics on larger timescales, as the self-gravity can require much smaller time steps in collapsing regions.

\subsection{Cloud Identification Algorithm}
\label{sec:cloud-id}

The cloud identification procedure implemented in our simulations has the following steps. All particles have a density calculated according to the SPH formalism. The first step in our scheme is to keep all particles above a certain density threshold ($\rho_{\mathrm{thresh}}$). Then in order to build a cloud catalogue from the remaining gas particles, we implement a Friends-of-Friends (FOF) \citep{HuchraGeller1982} algorithm, which is based on a linking length $l_{\mathrm{link}}$ parameter to relate particles within a given spatial scale. \citet{Dobbsetal2015} has tested a similar approach in cloud identification in SPH simulations. 

Additionally, we require that a clump has at least $50$ particles in order to keep it in our catalogue, which sets the minimum clump mass to $M = 31.25M_\odot$ given the particle mass in the simulation. We set $\rho_{\mathrm{thresh}} = 10 \,\mathrm{M}_\odot/\mathrm{pc}^3$, which in number density translates to $n \approx 311 \,\mathrm{cm}^{-3}$ (assuming $\mu = 1.299$ for the mean molecular weight). The linking length is set to $l_{\mathrm{link}} = 1$ pc. 

In this approach, the density cut determines the boundaries of a clump and the linking length implies that the algorithm can distinguish structures whose boundaries are separated by at least $1$ pc. For a given clump that it is separated by more than a $l_{\mathrm{link}}$ from other structures, the FOF algorithm will only link particles within the clump. Particles at the boundary will be able to link to other particles in the clump but it will not link anything outside it. However, if the edge of a different clump is within $l_{\mathrm{link}}$ of a particle in the edge of the first one, the algorithm will tag both as the same structure, which is a limitation of the FOF algorithm. In the case of not using a density cut, the linking length would still translate to a density cut because as the gas density decreases, the typical separation between SPH gas particles will increase to the point where it exceeds the linking length. 

In the present work, we are mainly interested in the large scale dynamics of dense gas clumps and not in particular properties that may be more sensitive to the criteria used to define a cloud. We use the identified clump's centre of mass positions and velocities to map their location and motion with respect to the galaxy.

\subsection{Kinematic Distance Method}
\label{sec:KD-method}

Given a cloud's line of sight velocity $V_{\mathrm{los}}$ and its Galactic longitude $l$, its galactocentric radius $R_{\mathrm{cloud}}$ and distance $D$ can be derived from the following equations, as described in \citet{Roman-Duvaletal2009}:

\begin{equation}
	R_{\mathrm{cloud}} = R_0 \sin l \left(\frac{V_c(R_{\mathrm{cloud}}) }{V_\odot \sin l + V_{\mathrm{los}}} \right),
	\label{eq:Rkin}
\end{equation}
where $R_0$ is the Sun's galactocentric position and $V_\odot$ is magnitude of the Sun's orbital velocity. This assumes that the cloud's motion follows a circular orbit with velocity $V_c(R_{\mathrm{cloud}})$. The distance is then given by
\begin{equation}
	D = R_0 \cos l \pm \sqrt{R_{\mathrm{cloud}}^2 - \left(R_0 \sin l \right)^2}.
	\label{eq:Dkin}
\end{equation}
The positive and negative roots are usually known as the ``far'' and ``near'' distances, respectively. The positive root should be used for a cloud located beyond the tangent point. The fact that the projection of a cloud's velocity on the line of sight can be the same for near and far points introduces the problem of the kinematic distance ambiguity (e.~g. \citealt{Roman-Duvaletal2009}). In a simulation we have the advantage of knowing \emph{a priori} the position of a cloud relative to the tangent point, so we can select which root to use.

\section{Cloud Kinematics and Streaming Motions}
\label{sec:cloud-kinematics}

\subsection{Cloud Rotation Curve and Streaming Motions}
\label{sec:gas-motions}

\begin{figure}
	\begin{center}
		\includegraphics[width=0.95\columnwidth]{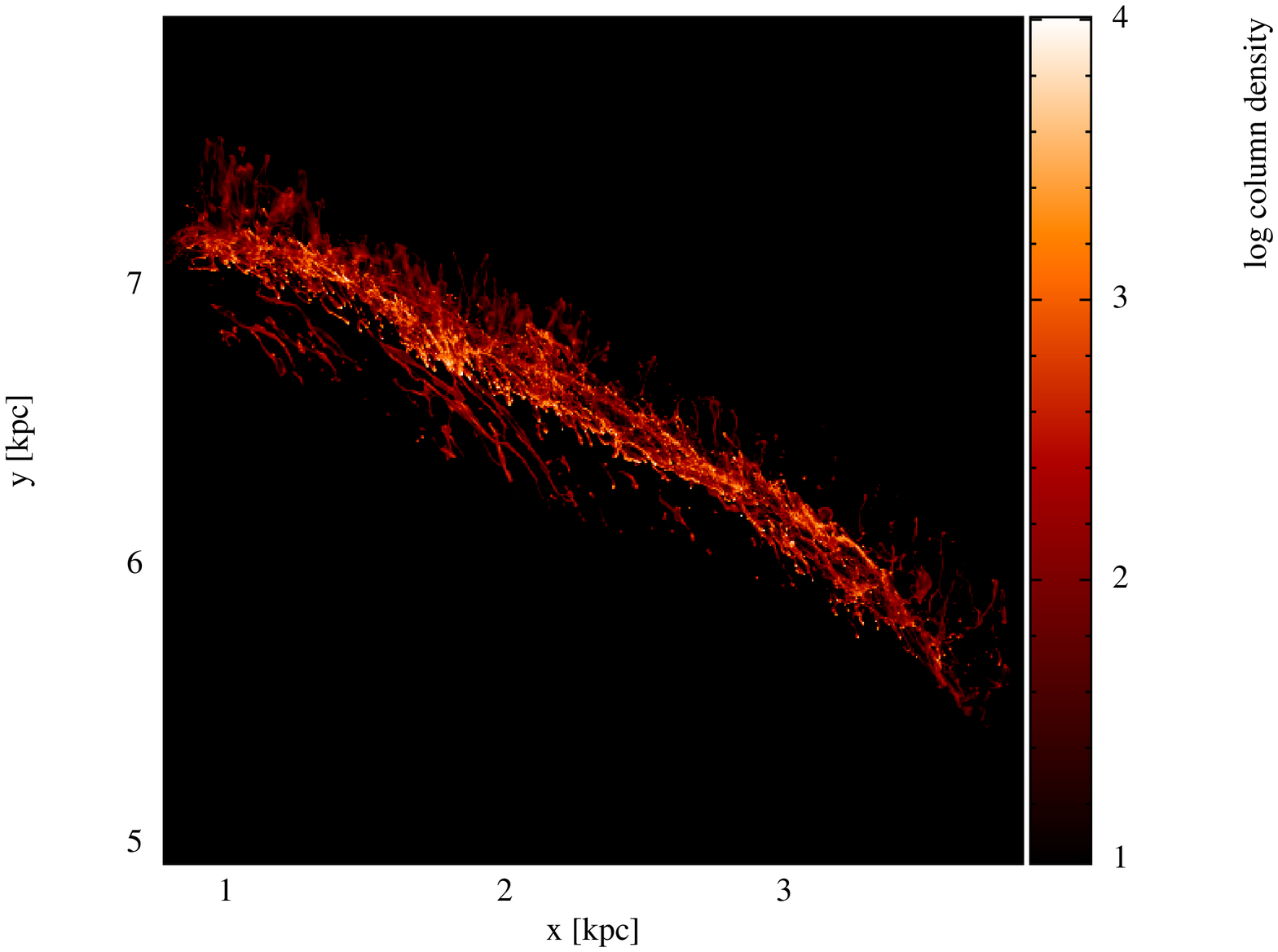}
		\includegraphics[width=0.95\columnwidth]{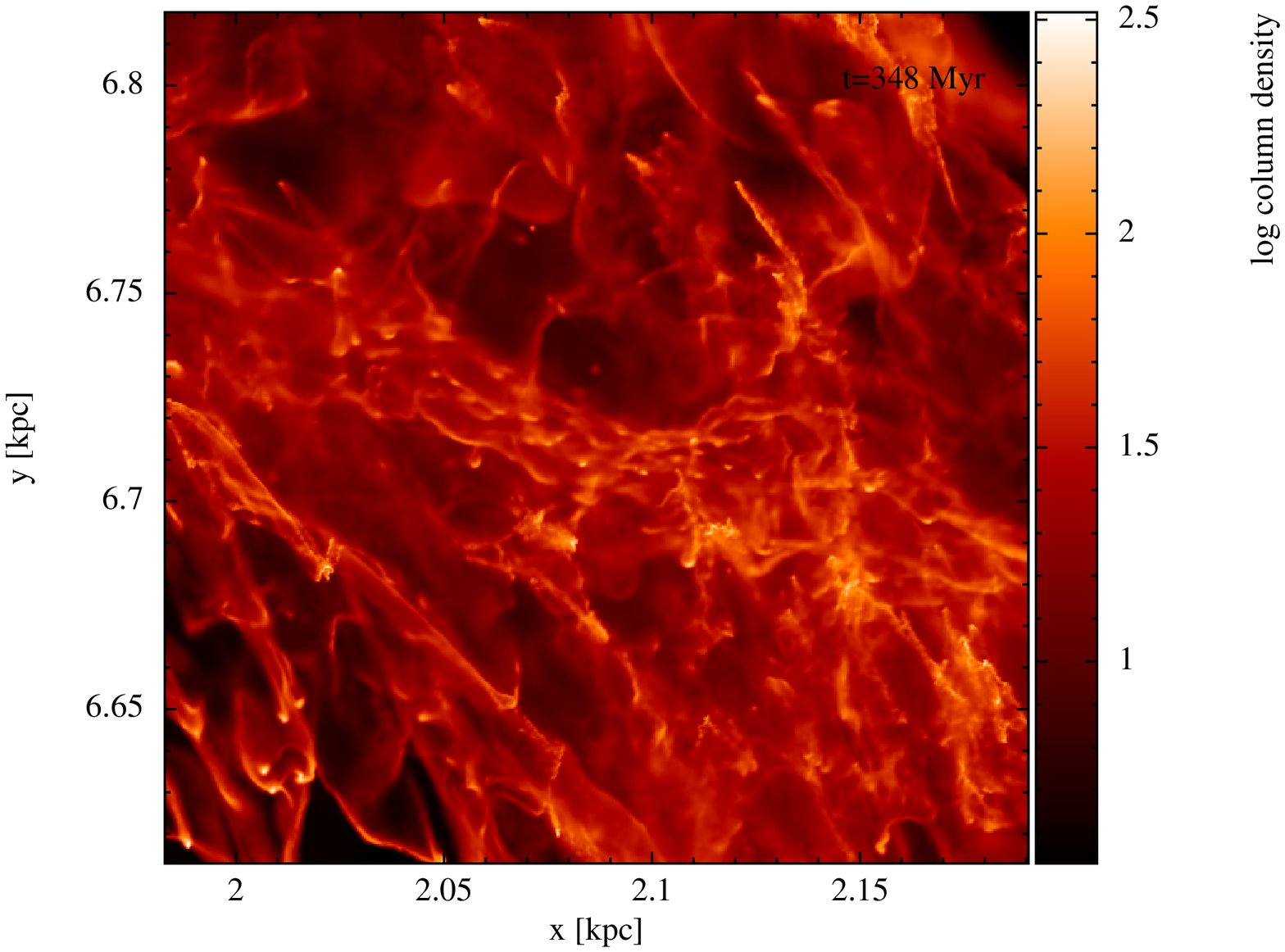}
		\includegraphics[width=0.95\columnwidth]{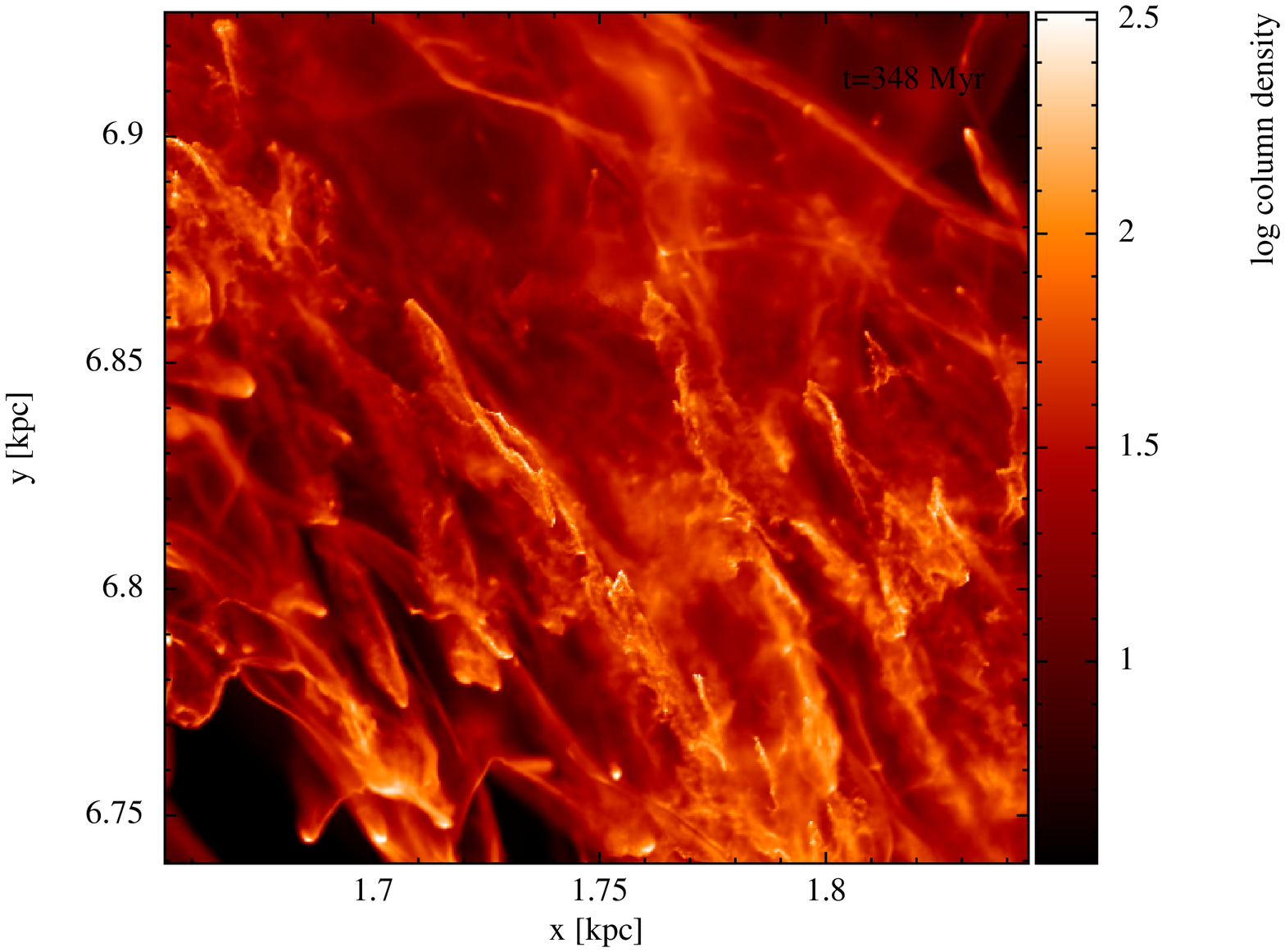}
		\caption{\emph{top panel}: gas surface density map of the spiral arm region; \emph{middle} and \emph{lower panel}: zoom-in snapshots showing in more detail the gas structures in the simulation.}
		\label{fig:gas-pics-1}
	\end{center}
\end{figure}

\begin{figure}
	\begin{center}
		\includegraphics[width=0.95\columnwidth]{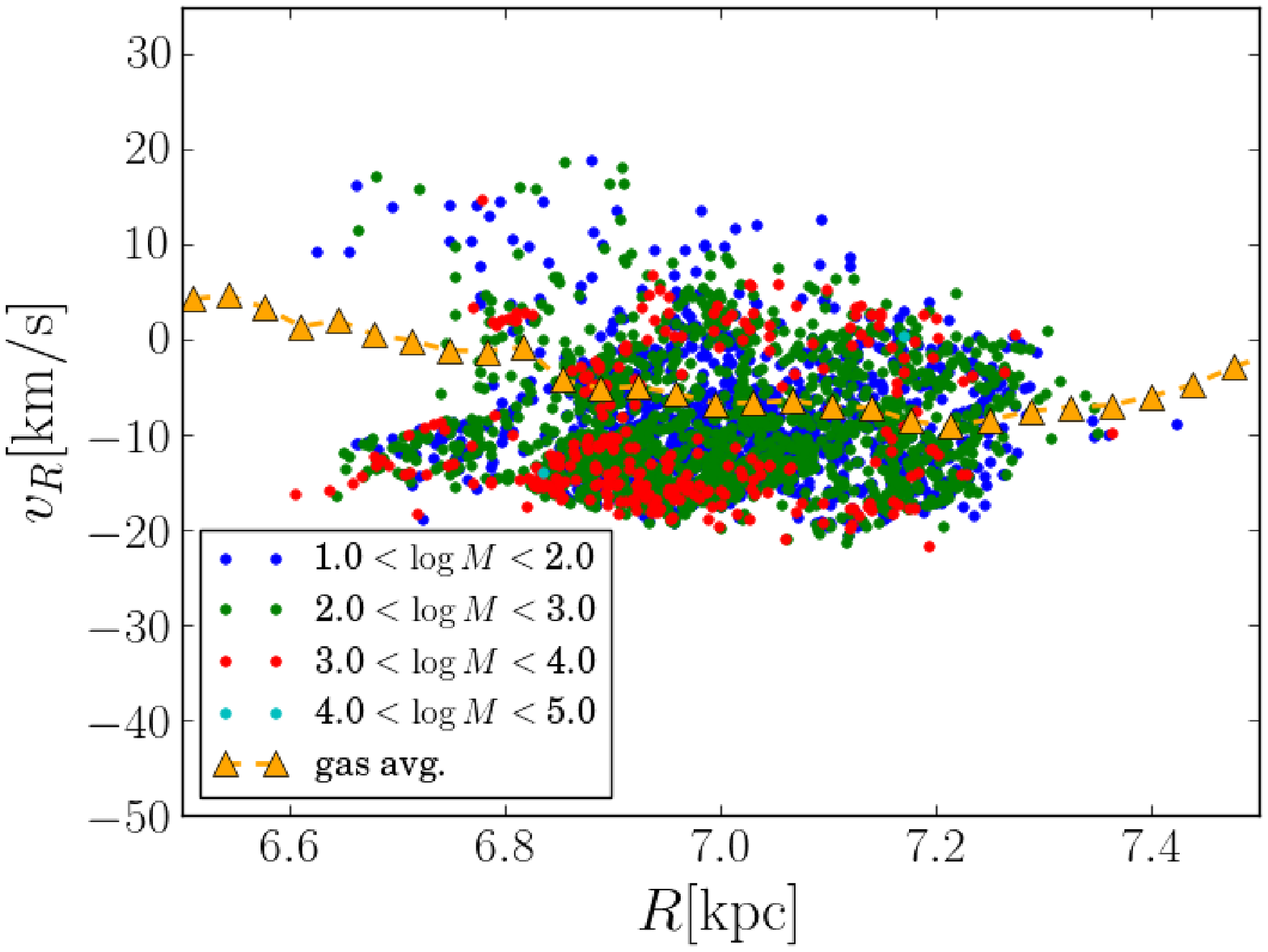}
		\includegraphics[width=0.95\columnwidth]{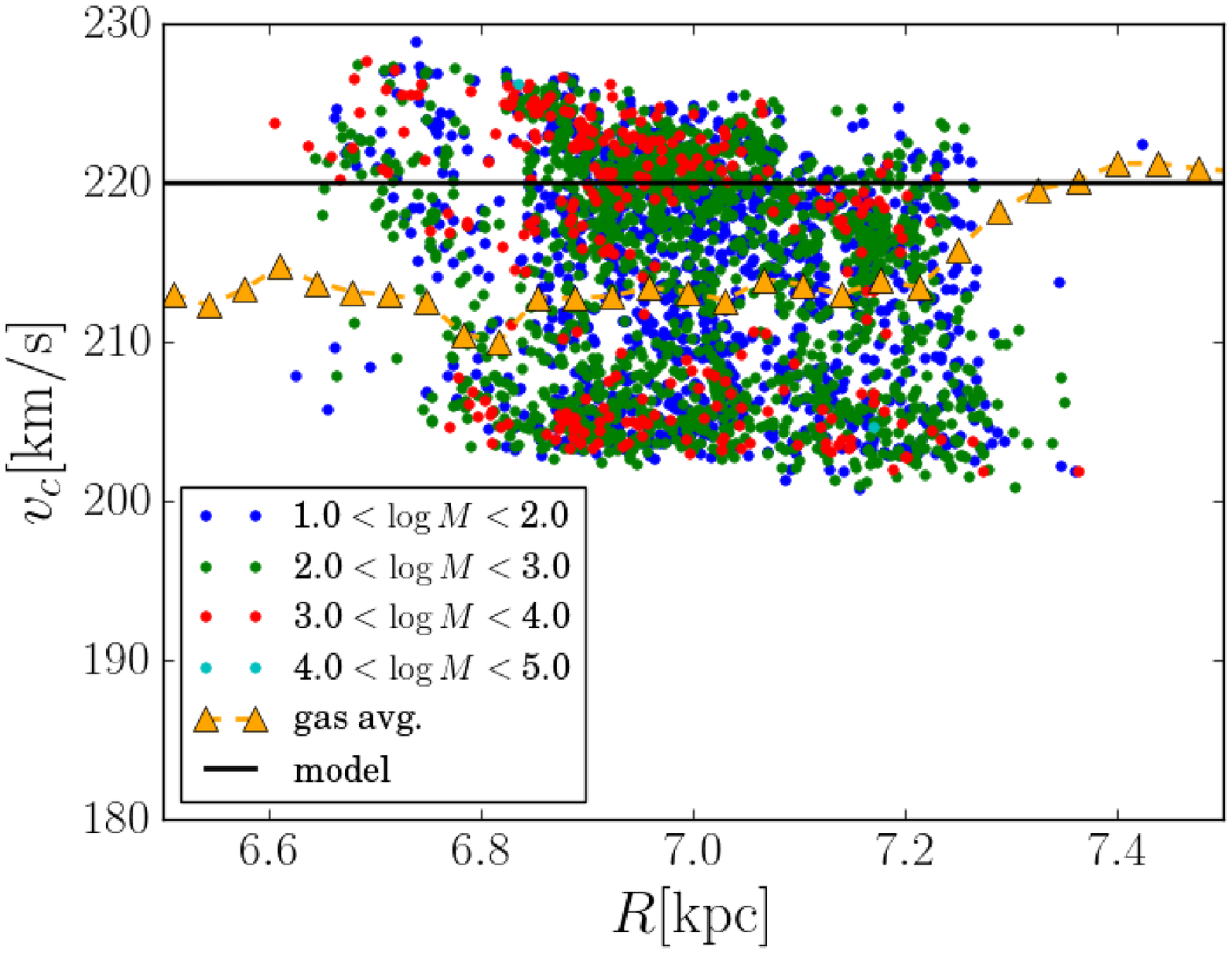}
		\caption{Cloud radial velocity (\emph{top panel}) and rotation curve (\emph{bottom panel}) colour coded by mass (in \msun~units). The orange triangles show the average velocity of the full gas distribution, which includes both cold and warm components.}
		\label{fig:arm-motions-SPAM760}
	\end{center}
\end{figure}

In Figure \ref{fig:gas-pics-1} we plot the spiral arm region in our simulation after $\approx 18.2$ Myr of evolution. We include two zoom-in snapshots to show in more detail the gas structures that form in our simulation. The mass of the identified clouds ranges from $31.2$ \msun~to $\sim 10^4$ \msun. The radial (galactocentric) $v_R$ and circular velocities $v_c$ as a function of radius obtained for this time are plotted in Figure \ref{fig:arm-motions-SPAM760}. These are the centre of mass velocities of the identified clumps. For comparison, the average velocity of all the gas (labelled as $\overline{v_R}$ and $\overline{v_c}$), including both cold and warm, binned by radius is also plotted.

\begin{figure}
	\begin{center}
		\includegraphics[width=0.95\columnwidth]{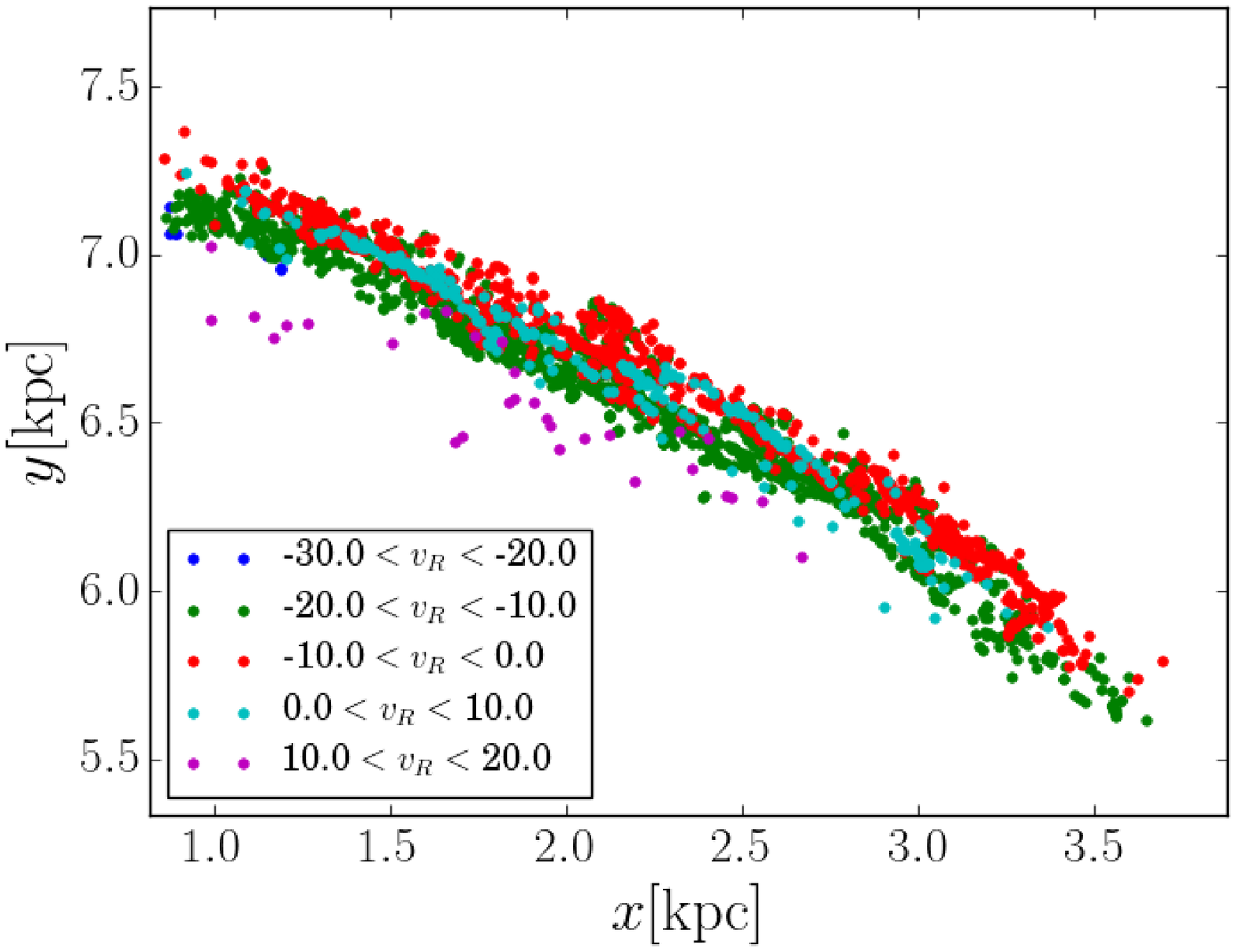}
		\includegraphics[width=0.95\columnwidth]{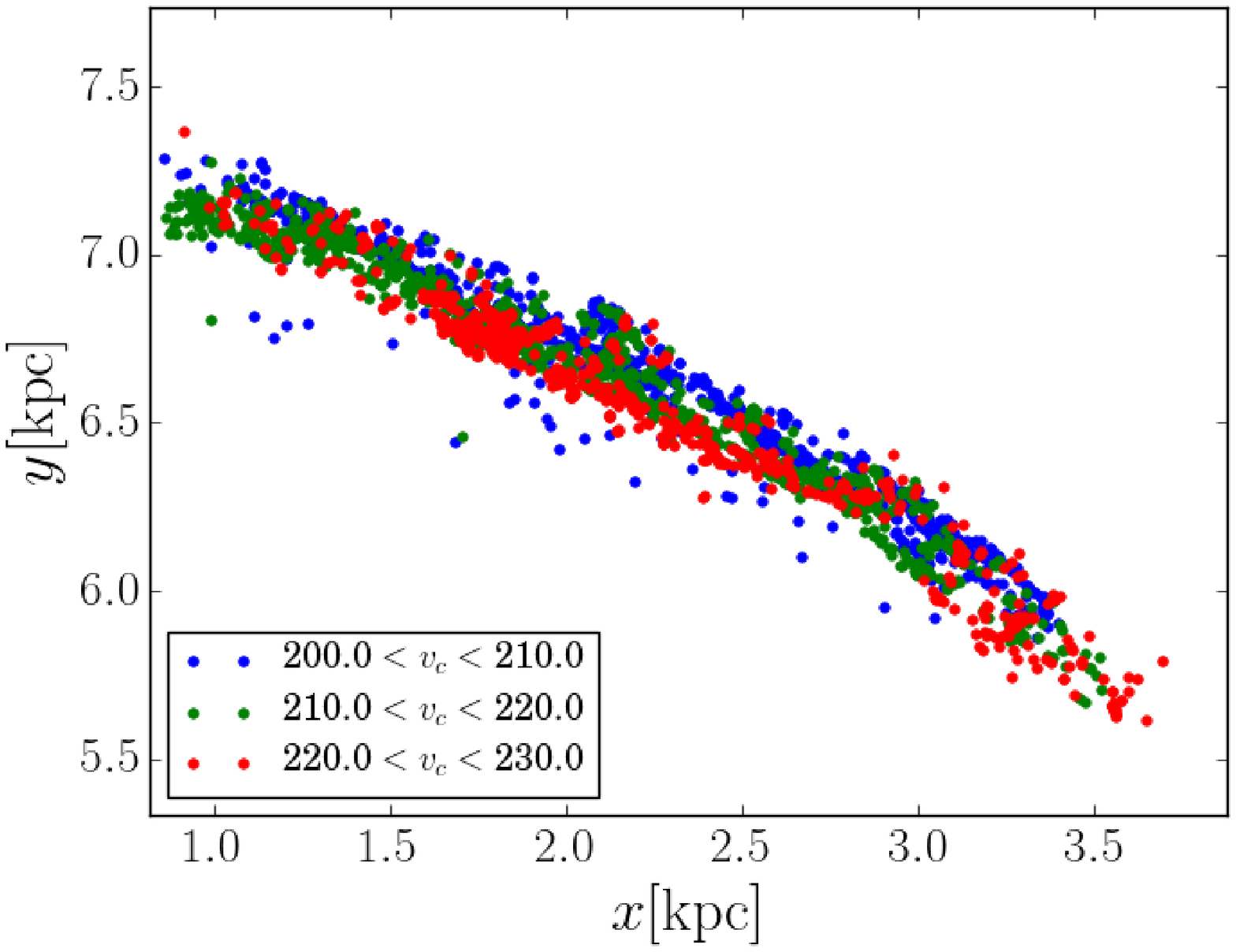}
		\caption{Cloud position map colour coded by their radial velocity (\emph{top panel}) and circular velocity (\emph{bottom panel}). The overall rotation of the galaxy is in a clockwise sense. Clouds on the left hand side of the distribution have a higher circular velocity than those on the right side.}
		\label{fig:cloud-vcirc-xymap}
	\end{center}
\end{figure}

In the radial component, the average radial motion of the gas $\overline{v_R}$ does show a significant shift from $v_R = 0$ (orange triangles in Figure \ref{fig:arm-motions-SPAM760}, upper panel). This radial component has the largest deviation at $R = 7.2$ kpc, where $\overline{v_R} = -9$ \kms. The clouds' radial motions show a scatter from $-21.7$ to $18.9$ \kms, with an average $v_R = -9.3$ \kms. However, when their motions are compared to the average motion of the gas $(\delta v_R(\mathrm{cloud}) = v_R(\mathrm{cloud}) - \overline{v_R}(\mathrm{gas}))$, we find a scatter from $-18.4$ \kms~to $23.9$ \kms. The average of this difference is $\overline{\delta v_R}(\mathrm{cloud}) = -3.4$ \kms, showing the cloud velocity average is slightly shifted from the overall radial gas motions. The cloud-to-cloud radial velocity dispersion obtained is $\sigma_R = 6.4$ \kms. Figure \ref{fig:cloud-vcirc-xymap} (top panel) shows the radial velocity component as a function of the cloud position. Most of the clouds are moving with negative $v_R$, though there are some clouds near the centre of the distribution moving with positive values.

In the azimuthal component (see Figure \ref{fig:arm-motions-SPAM760}, lower panel), there is a large scatter in the cloud's velocities. However, on a closer inspection, it is possible to see two main groups: one with $v_c$ around $220$ \kms and another with $v_c$ around $205$ \kms. The average azimuthal velocity of the gas $\overline{v_c}$ passes between the two groups. Figure \ref{fig:cloud-vcirc-xymap} shows that the clouds of the fast group are on the upstream side of the arm where gas is entering the region and the clouds in the slower group are on the opposite side. The average gas velocity and most of the cloud's velocities tend to be slower than the rotation curve.

The clouds' circular velocity is scattered between $200.8$ \kms and $228.9$ \kms with a mean $v_c = 214.2$ \kms. When compared to the average gas velocity $(\delta v_c(\mathrm{cloud}) = v_c(\mathrm{cloud}) - \overline{v_c}(\mathrm{gas}))$, the distribution is scattered between $-16.5$ \kms~and $18.3$ \kms, with the average difference being $\overline{\delta v_c}(\mathrm{cloud}) = -1.1$ \kms. This indicates that the cloud distribution as a whole does not have a significant net difference with respect to the overall gas motion, but as shown in Figure \ref{fig:arm-motions-SPAM760}, there is a cloud group rotating faster and a second one rotating slower than the average gas velocity. 

In this component, the velocity dispersion is $\sigma_\phi = 6.9$ \kms. Taking into account both components $(v_x$ and $v_y$) on the mid-plane of the galaxy, the cloud-to-cloud velocity dispersion of the magnitude of the velocity on the plane gives $\sigma_v = 7.1$ \kms.

\begin{figure}
	\begin{center}
		\includegraphics[width=0.95\columnwidth]{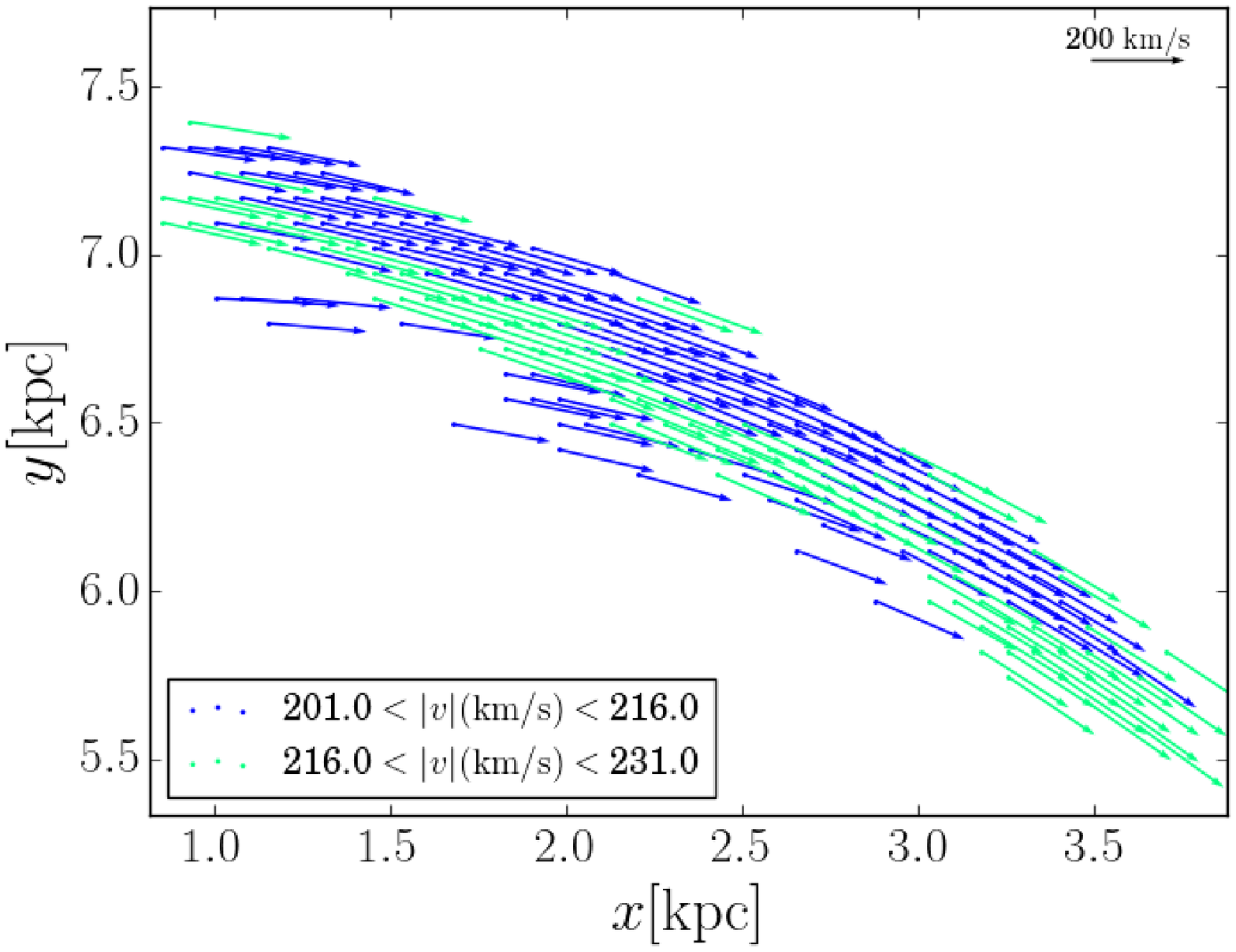}
		\includegraphics[width=0.95\columnwidth]{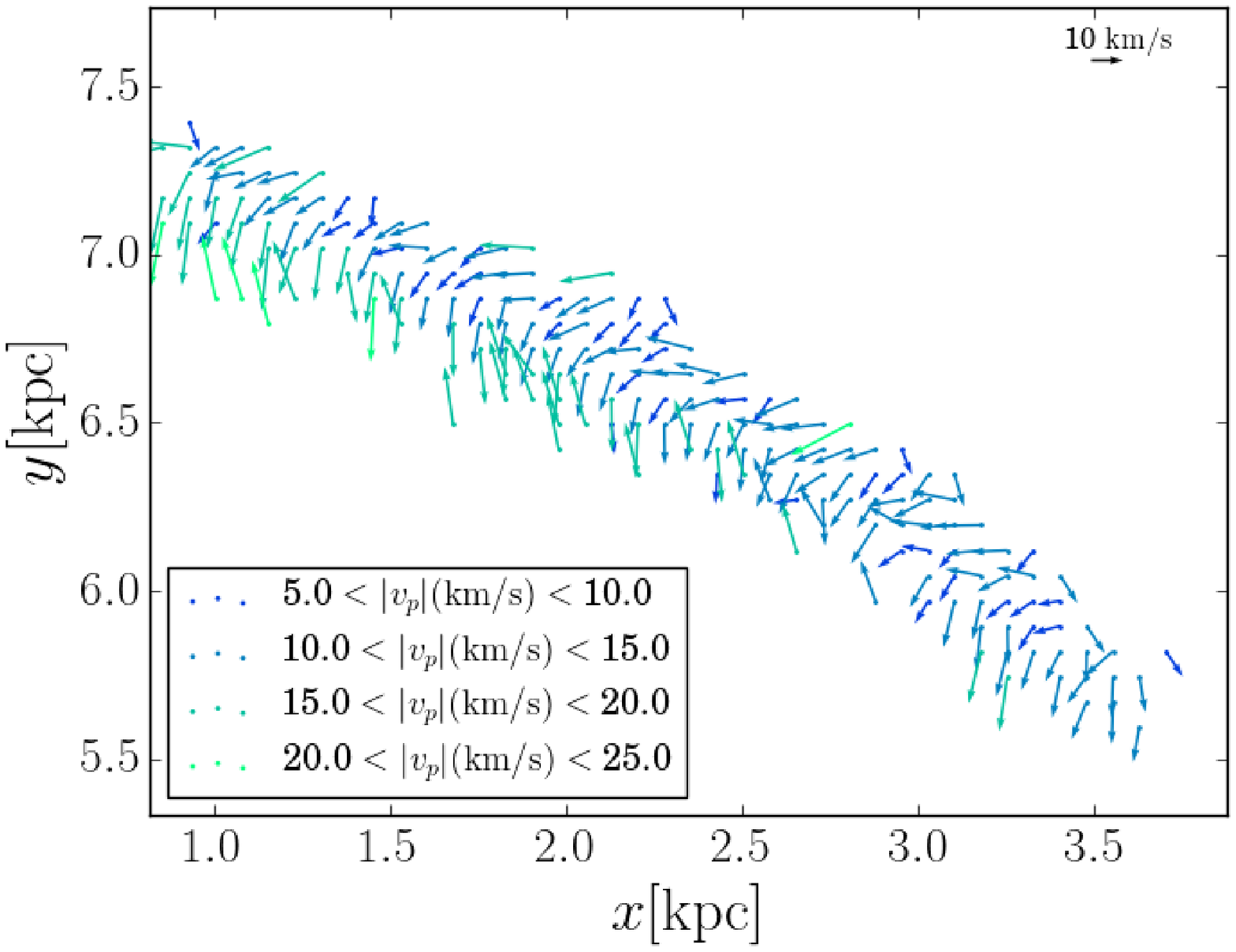}
		\caption{Cloud velocity field (\emph{top panel}) and peculiar velocity field (\emph{bottom panel}) averaged in cells with size length of $75$ pc.}
		\label{fig:cloud-vel-fields}
	\end{center}
\end{figure}

Peculiar velocities, as compared to the local circular velocity, can be a more useful diagnostic for comparing with observations. We obtain the peculiar motion by subtracting the circular velocity at the cloud's position from the actual velocity. Figure \ref{fig:cloud-vel-fields} shows the total (top panel) and peculiar (bottom panel) velocity fields averaged in bins of $75$ pc using only the velocity on the galaxy plane.

The peculiar motions also show a net component toward the galactic centre due to the fact that the cloud distribution has a net radial motion in this direction.

The peculiar velocities show a net motion that appears to travel inwards through the spiral arm, leaving from the inner, trailing edge. This is actually in the opposite direction of the cloud motions that enter the spiral arm from the trailing edge and leave through the leading side. This is due to the pattern speed of the spiral potential, which lags the circular velocity by nearly 30 \kms~at the Solar Circle.

\begin{figure}
	\begin{center}
		\includegraphics[width=0.95\columnwidth]{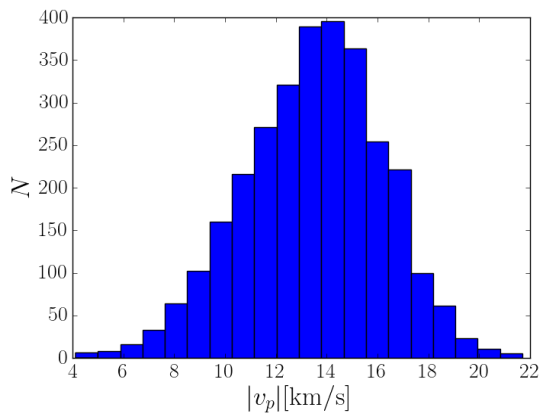}
		\caption{Histogram of the peculiar velocity vector magnitude $|v_p|$. Typical motions are $\approx 15$ \kms and are not larger than about $10 \%$ of the rotation curve.}
		\label{fig:cloud-velpec-hist}
	\end{center}
\end{figure}

Figure \ref{fig:cloud-velpec-hist} shows a histogram of the magnitude of the peculiar motion $|v_p|$, which has typical values of $\approx 15$ \kms.

\section{Kinematic Distance Errors}
\label{sec:KD-errors}

\subsection{Distance Errors due to a Cloud-to-Cloud Velocity Dispersion}
\label{sec:effect-vdisp-kd}

To assess how a cloud-to-cloud velocity dispersion in molecular clouds propagates into an error in the kinematic distance, we created a mock cloud catalogue by assigning to each cloud the local axisymmetric circular velocity. First, we test the cases where a radial component is added from a normal distribution with dispersions: $\sigma_R = 1.0,\, 5.0,$ and $10$ \kms. We repeat the process including an azimuthal dispersion $\sigma_\phi$ with the same values. We study the velocity components independently in order to explore the error introduced in each component, which will help to interpret the results from the actual kinematics in the simulation.

\begin{figure}
	\begin{center}
		\includegraphics[width=0.99\columnwidth]{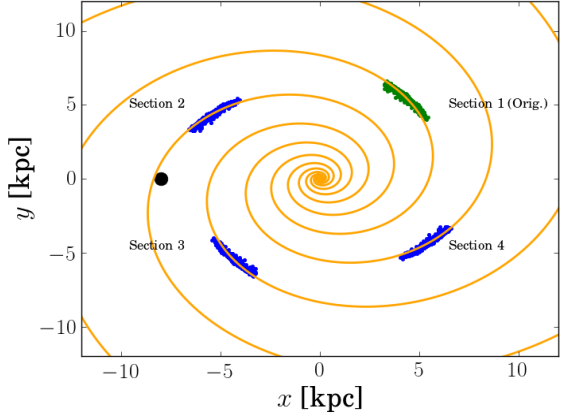}
		\caption{The clouds identified in the original simulation are in section I (upper right corner). The rest are copies placed in the equivalent positions on the remaining spiral arms in order to apply the kinematic distance method in all the sections. The orange curves trace the spiral arms to which the clouds are associated. The solid black circle is the observing point.}
		\label{fig:mock-positions-1}
	\end{center}
\end{figure}

To estimate the distance, we take an observing point at $\boldsymbol{R}_0 = \langle-8.0, 0.0, 0.0\rangle$ kpc assuming, for simplicity, that it moves at its local circular velocity: $v_c(R_0) = 220$ \kms. Our Milky Way model has 4 arms, so we copied the positions and velocities of clouds identified in the simulated region and placed them in the corresponding position on the other arms, as shown in Figure \ref{fig:mock-positions-1}. This allows us to test the kinematic distance method in the 1st and 4th galactic quadrants. For convenience in presenting our results, we define the region on the upper right from the origin as section 1 and the section number increases in the counter-clockwise direction.

\begin{figure}
	\begin{center}
		\includegraphics[width=0.95\columnwidth]{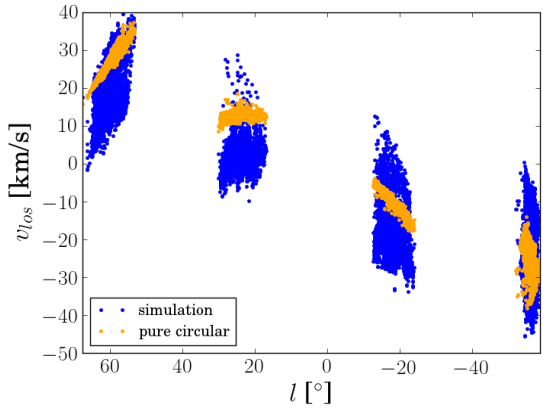}
		\caption{Line of sight $v_\mathrm{los}$ vs. galactic longitude $l$ map comparing the actual kinematics of the clouds in the simulation with the values that would result from a purely circular velocity at the cloud's position. The actual $v_\mathrm{los}$ is not necessarily symmetric around the value expected from the circular velocity.}
		\label{fig:lv-plot}
	\end{center}
\end{figure}

Figure \ref{fig:lv-plot} shows the line of sight velocity $v_{\mathrm{los}}$ vs. galactic longitude $l$ for the cloud distribution in each spiral arm; $v_{\mathrm{los}}$ is obtained by projecting the cloud's velocity relative to the observing point on the line of sight. The orange dots in Figure \ref{fig:lv-plot} show the values for clouds moving in circular orbits. The small spread is caused by the spread in positions of the clouds. The blue dots show the line-of-sight velocities obtained from the clouds' actual motions in the simulation. This shows that $v_{\mathrm{los}}$ is not necessarily symmetric around the values expected from purely circular motions. The deviations can be larger than $10$ \kms.

\begin{figure*}
	\includegraphics[width=0.95\columnwidth]{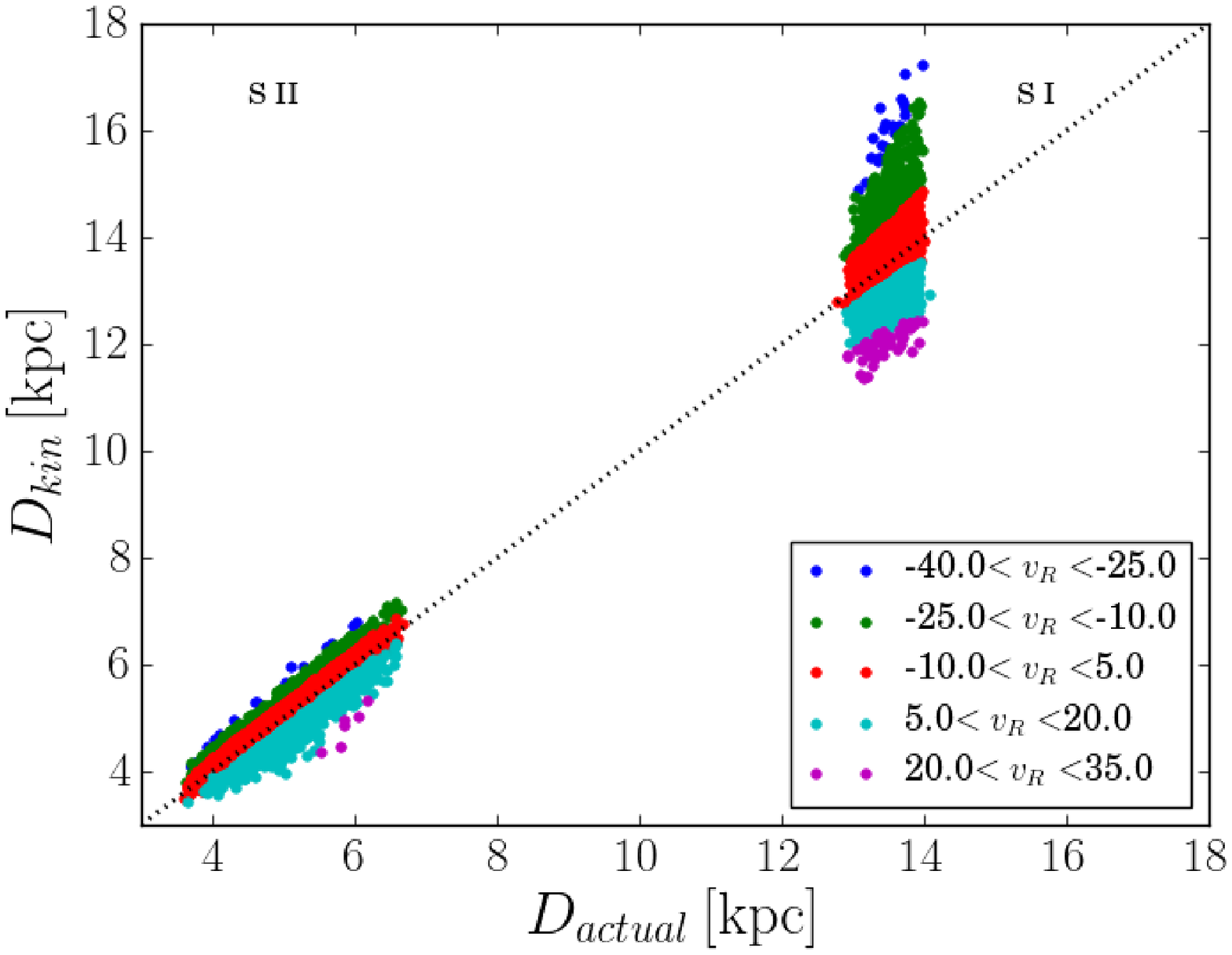}
	\includegraphics[width=0.95\columnwidth]{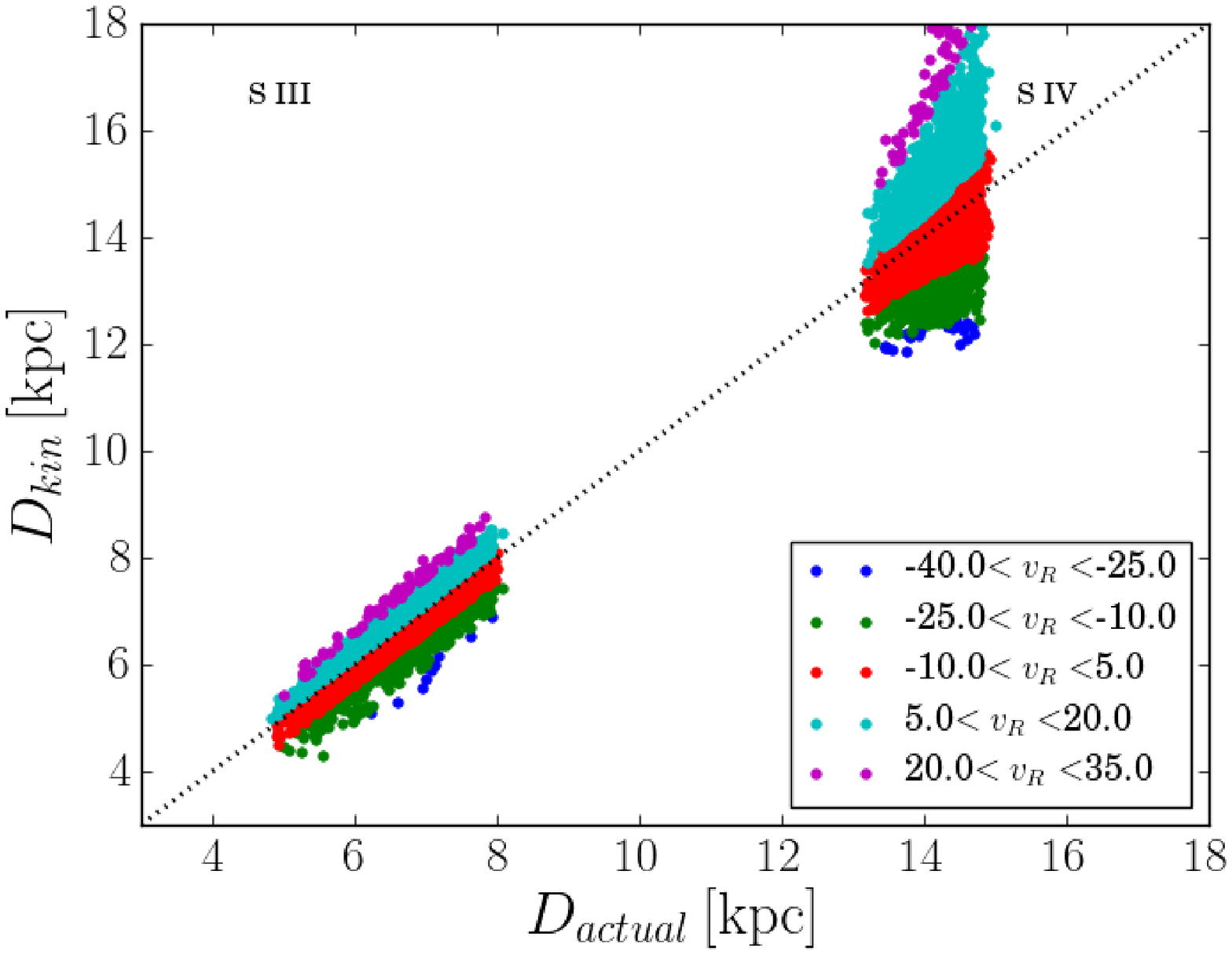}
	\caption{Cloud kinematic distance \dkin~compared to the actual value \dact~for a radial velocity scatter around the rotation curve of $\sigma_R = 10$ \kms. Sections I and II are shown in the \emph{left panel}. Sections III and IV are shown in the \emph{right panel}. Radial streaming components affect the distance estimate for more distant clouds.}
	\label{fig:kdist-DISP760-rad-disp10p0}
\end{figure*}

\begin{figure*}
	\includegraphics[width=0.95\columnwidth]{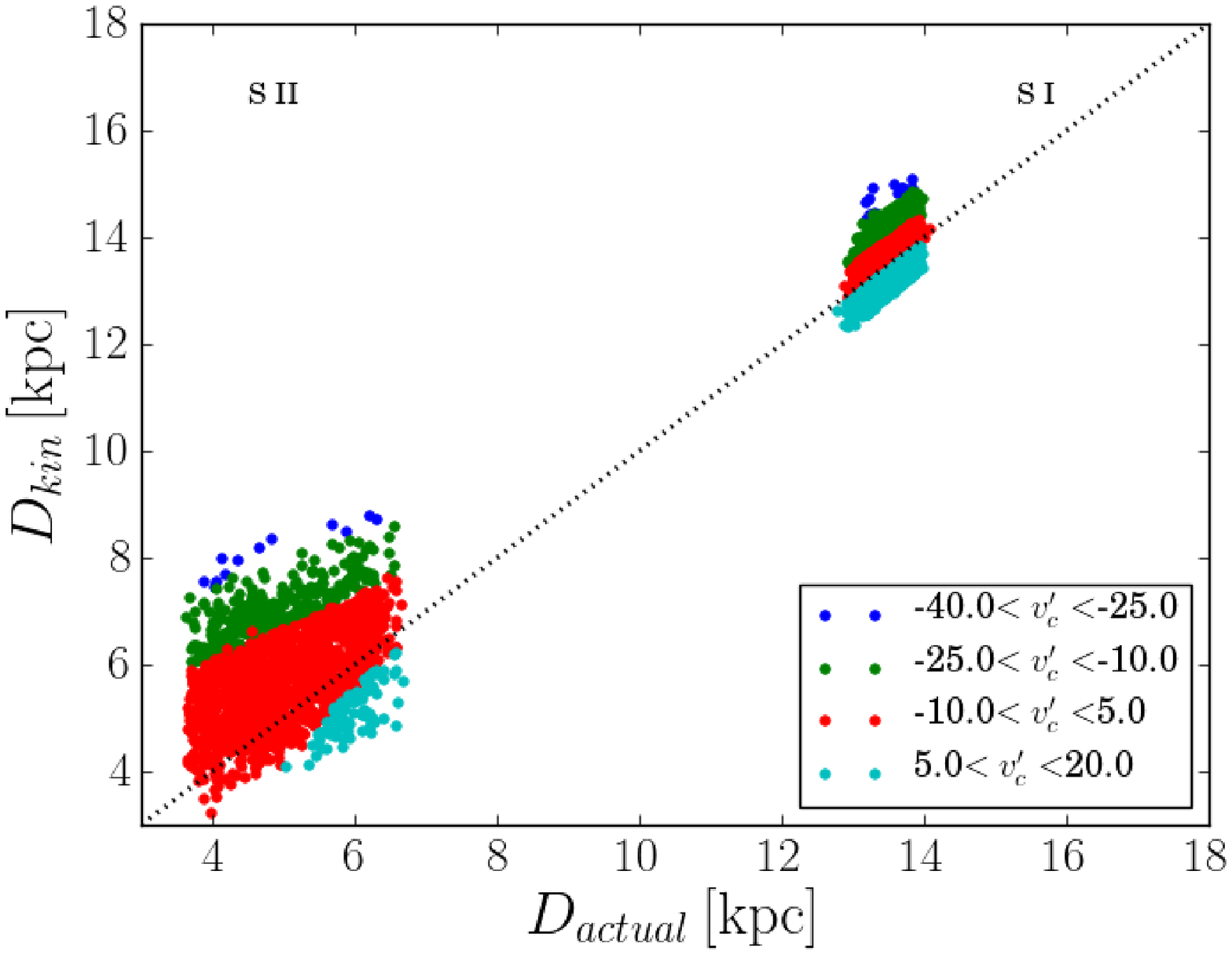}
	\includegraphics[width=0.95\columnwidth]{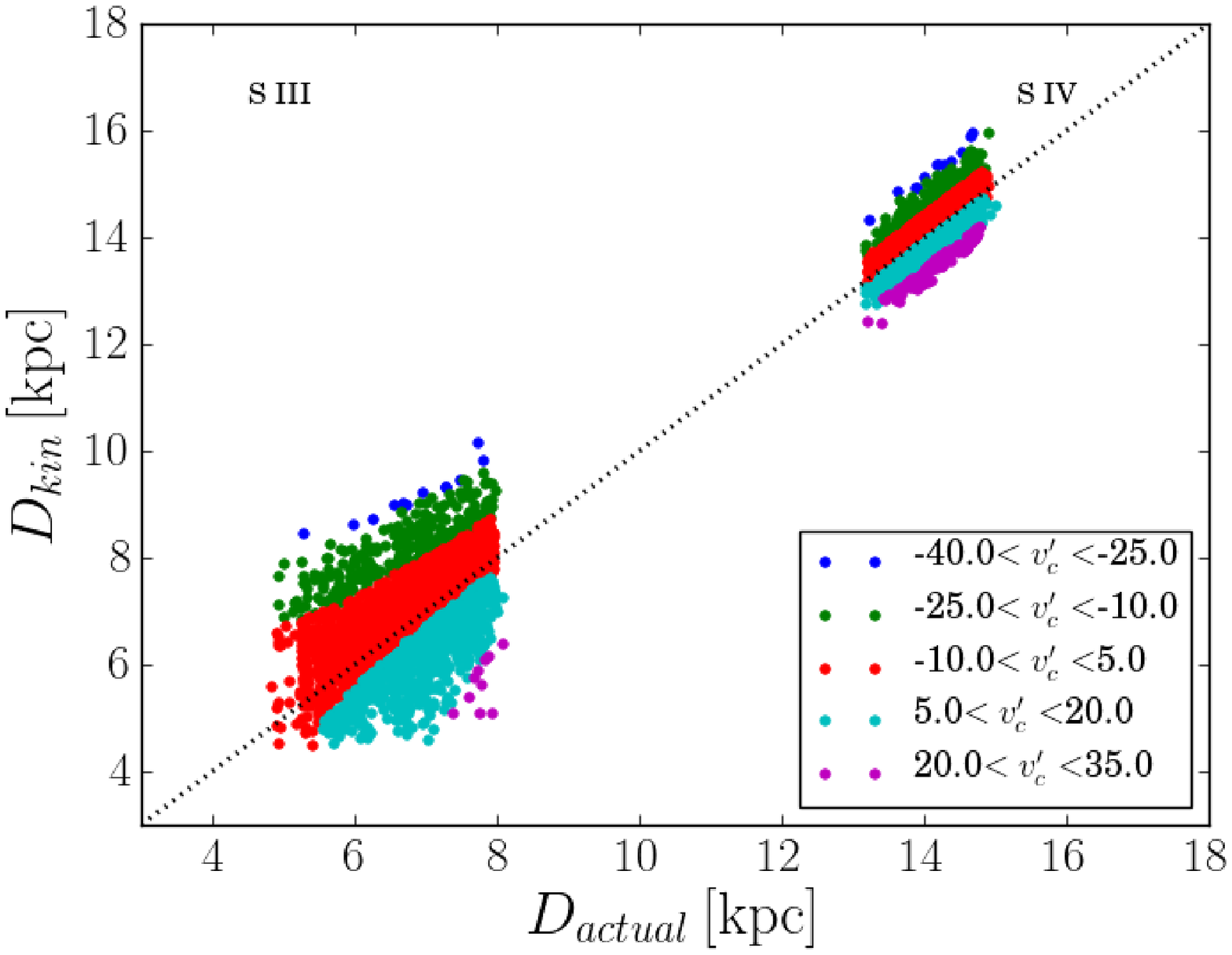}
	\caption{Cloud distance derived from the kinematic method \dkin~compared to the actual value \dact~for an azimuthal scatter around the rotation curve of $\sigma_\phi = 10$ \kms. Sections I and II are shown in the \emph{left panel}. Sections III and IV are shown in the \emph{right panel}. Azimuthal streaming components affect the distance estimate for nearby clouds.}
	\label{fig:kdist-DISP760-azim-disp10p0}
\end{figure*}

When a velocity scatter is added in the radial direction, the error is largest in sections I and IV (see  Figure \ref{fig:kdist-DISP760-rad-disp10p0}). The distance can be overestimated by as much as $4$ kpc for some clouds. The effect of adding a velocity scatter in the azimuthal direction is very systematic. The results are shown in Figure \ref{fig:kdist-DISP760-azim-disp10p0}, where the plots are colour-coded by the cloud's relative circular velocity: $v_c' = v_c(\mathrm{cloud}) - v_c$. For clouds moving slower than the rotation curve $(v_c' < 0)$, the distance is overestimated. The effect is reversed for the faster moving clouds $(v_c' > 0)$. 

\begin{figure}
	\includegraphics[width=0.95\columnwidth]{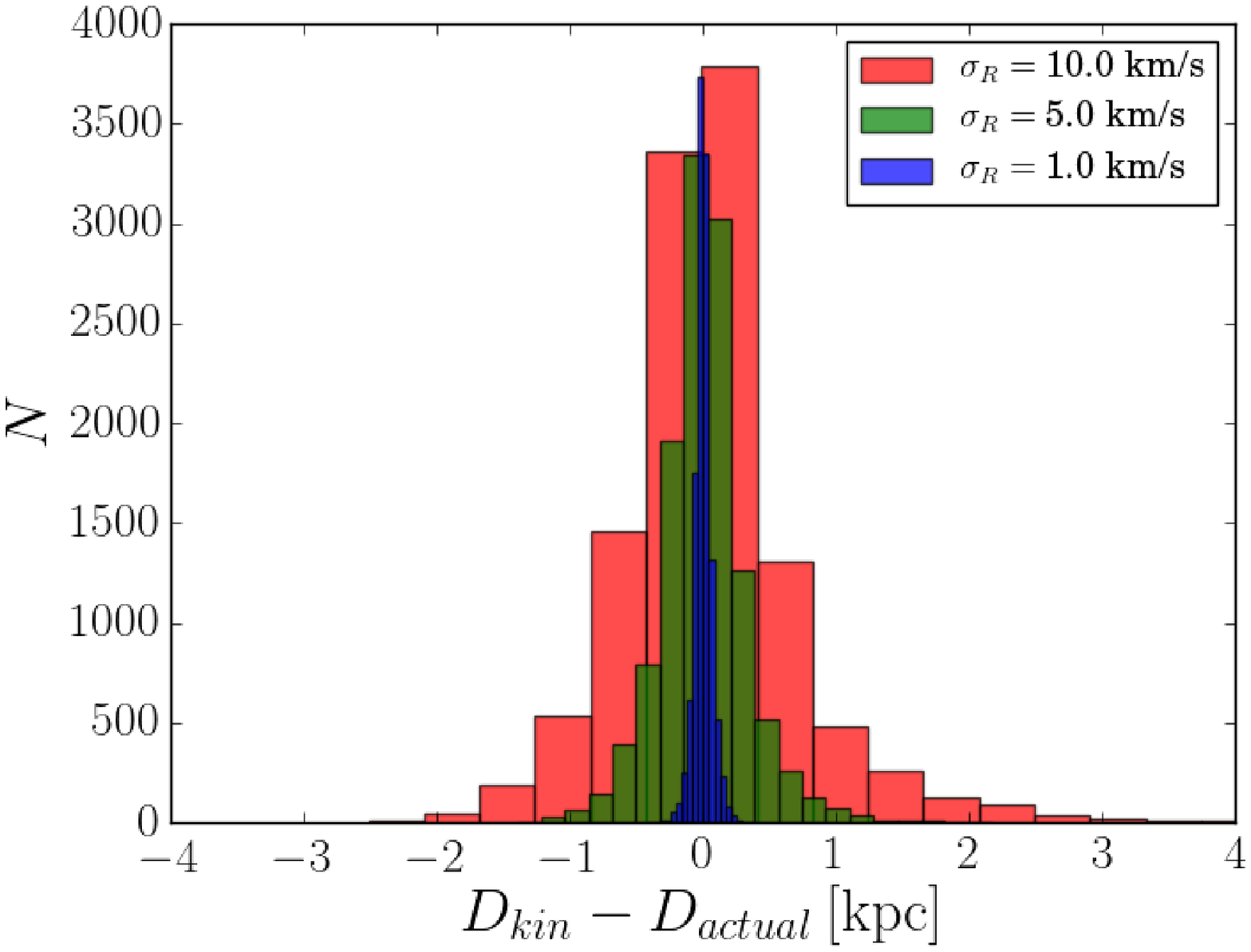}
	\includegraphics[width=0.95\columnwidth]{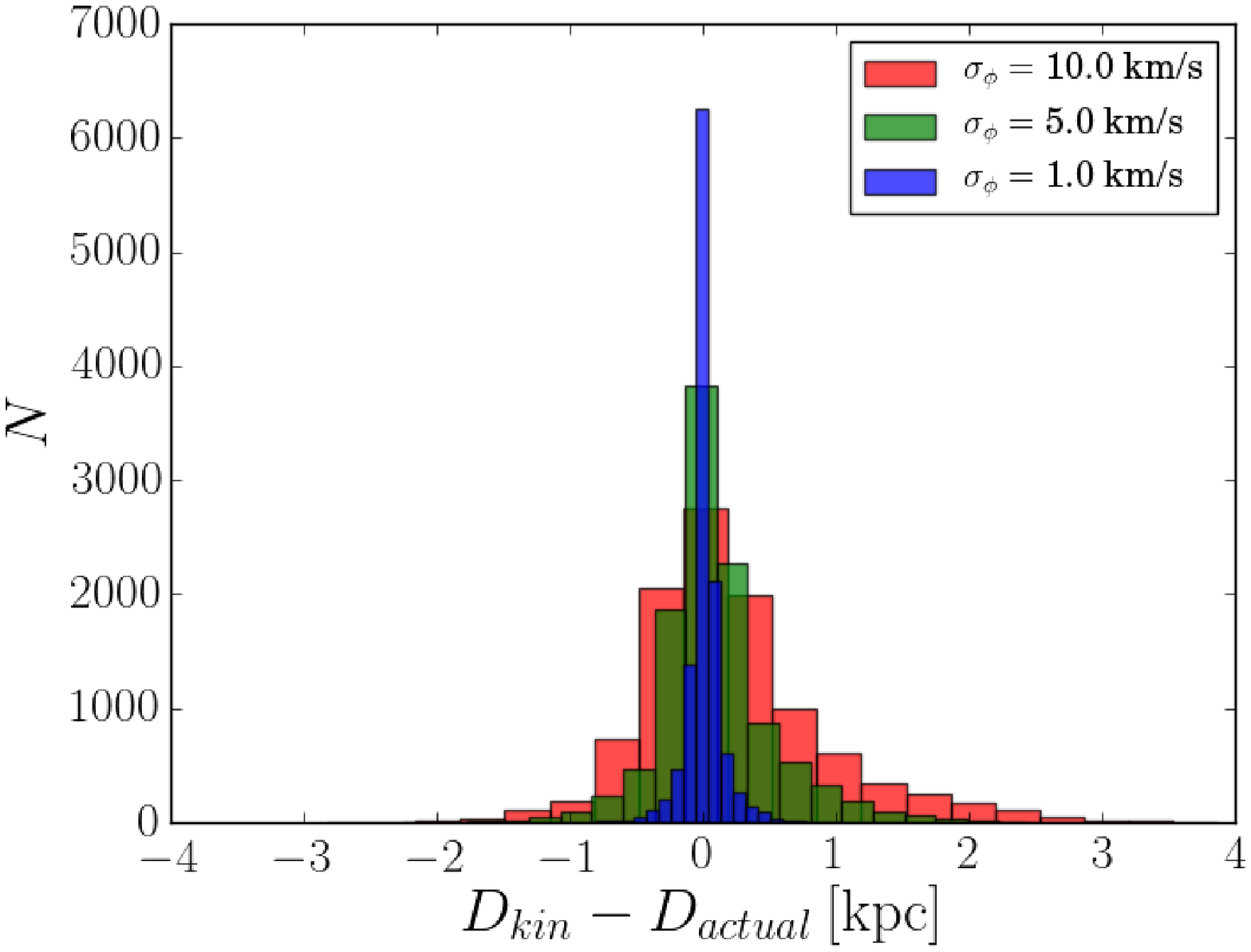}
	\caption{Distance error distributions for the case of a radial velocity scatter $\sigma_R$ (\emph{top panel}) and an azimuthal velocity scatter $\sigma_\phi$ (\emph{lower panel}) in the cloud velocities with respect to the circular velocity.}
	\label{fig:kindist-error-histograms}
\end{figure}

The distance error ($D_{\mathrm{kin}} - D_{\mathrm{actual}}$) distribution is given in Figure \ref{fig:kindist-error-histograms}. The upper panel shows the results from a radial velocity scatter. For $\sigma_R = 1.0$ \kms, the standard deviation of the error distribution is $0.06$ kpc and increases up to $0.673$ kpc for $\sigma_R = 10$ \kms. For the latter the distance error ranges between $-2.51$ kpc and $5.83$ kpc. Although this is large, $88\%$ of the clouds have errors within 1 kpc. The lower panel shows the error distribution for the azimuthal velocity scatter. For $\sigma_\phi = 1.0$ \kms, the standard deviation is $0.12$ kpc. For $\sigma_\phi = 10$ \kms, the deviation is $0.70$ kpc with the errors ranging from $-2.83$ kpc to $3.87$ kpc.

\begin{figure}
	\begin{center}
		\includegraphics[width=0.99\columnwidth]{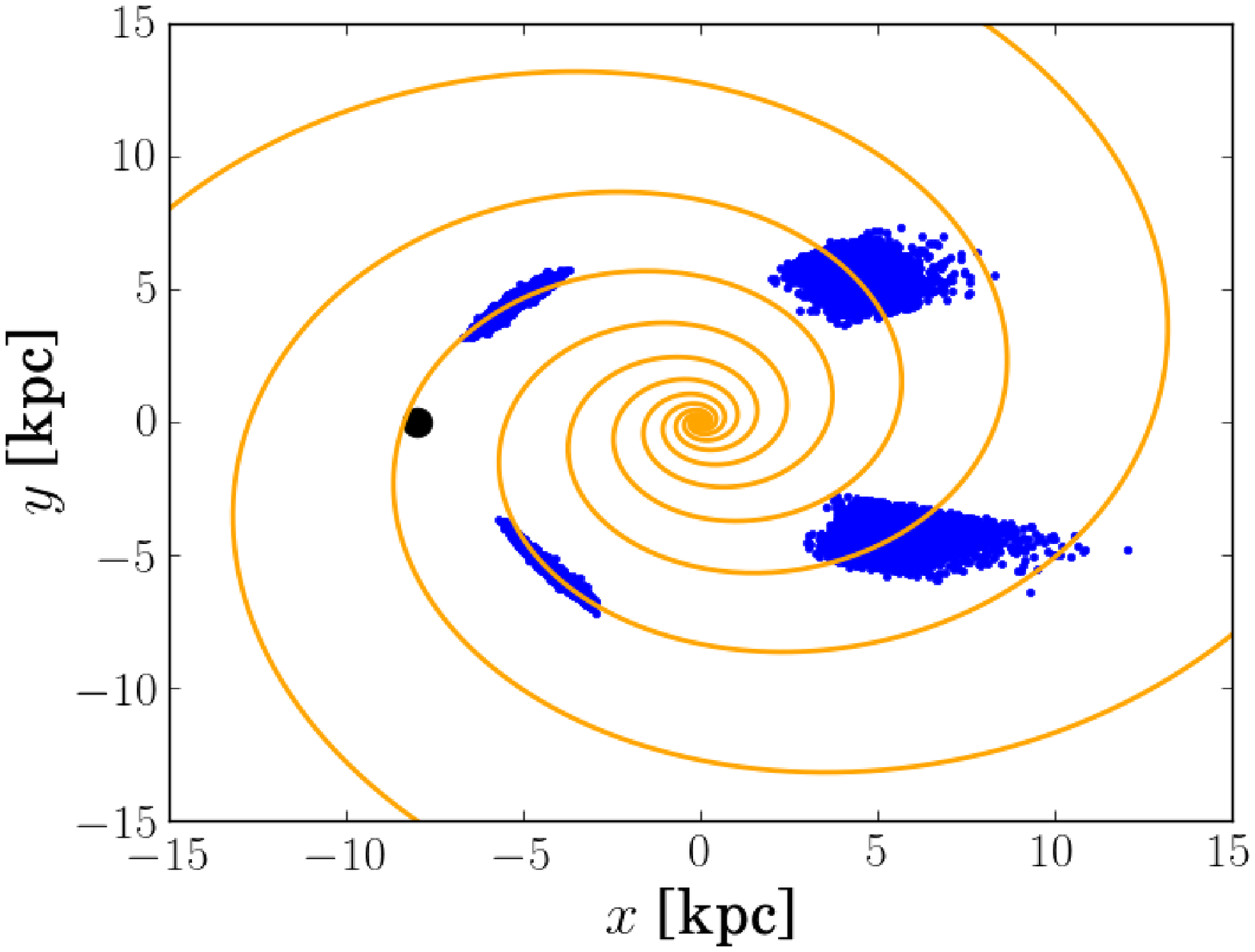}
		\includegraphics[width=0.99\columnwidth]{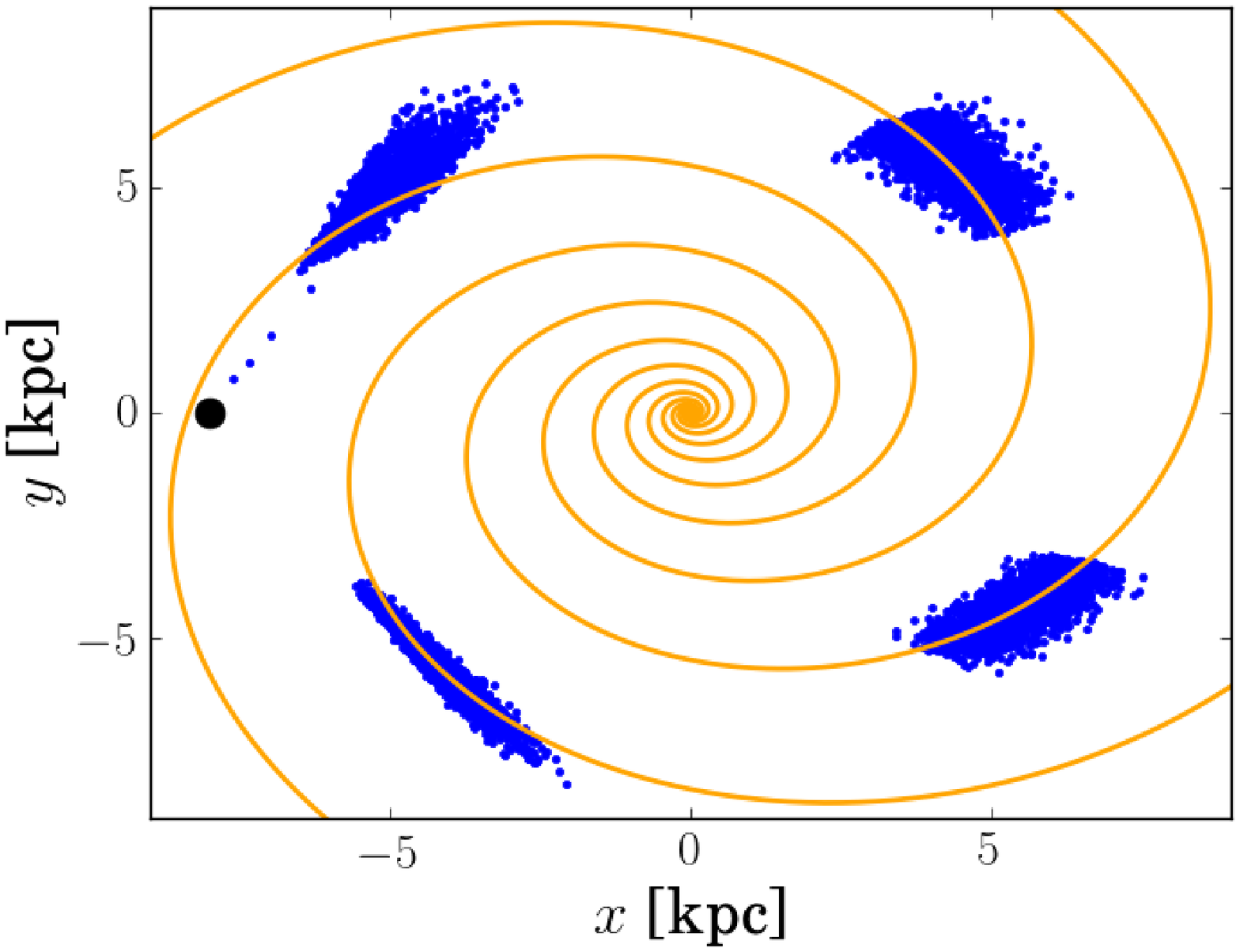}
		\caption{Cloud positions recovered from the kinematic distance. The clouds were given the circular velocity at their galactocentric radius plus a component from a cloud-to-cloud velocity dispersion $\sigma$ symmetric around the rotation curve. No net streaming components where added. The \emph{top panel} shows the case for $\sigma_R = 10$ \kms and the \emph{bottom panel} the case for $\sigma_\phi = 10$ \kms.}
		\label{fig:recovered-pos-disp10p0}
	\end{center}
\end{figure}

Figure \ref{fig:recovered-pos-disp10p0} shows the positions recovered from kinematic distances. The upper and lower panels show the results for radial and azimuthal cloud-to-cloud velocity dispersion, respectively. Both correspond to a $10$ \kms~dispersion. The recovered cloud distribution appears rather distorted in all sections when compared to the original positions (see Figure \ref{fig:mock-positions-1}). 

The upper panel of Figure \ref{fig:recovered-pos-disp10p0} shows that clouds with the largest distance errors are in sections I and IV. This is a consequence of the fact that they lie on a line where the contribution of the radial velocity to the line of sight projection is more important. The cloud groups in sections II and III appear to be less affected by a radial streaming component. The lower panel of Figure \ref{fig:recovered-pos-disp10p0} shows the recovered positions for the case when the velocity scatter was applied in the azimuthal direction. The results for all sections show a significant scatter. The cloud groups in sections II and III show the highest errors because they lie closer to the tangent line, where the azimuthal component dominates the line of sight velocity.

This simple model already shows that a cloud-to-cloud dispersion introduces large errors in the kinematic distance estimate. For the worst case tests, the spread in the error distribution is $\approx 0.6$ kpc. This is not negligible considering that an arm may be a few $100$ pc wide. We find that a radial streaming component can introduce important errors and that clouds with an azimuthal velocity faster (or slower) than the circular velocity will have their distance underestimated (or overestimated). This is a problem when determining cloud positions near spiral arms. Our simulation shows that clouds in the near side of the arm with respect to the observer are moving faster than those on the far side (see Figure \ref{fig:cloud-vcirc-xymap}). The kinematic estimate would underestimate distances to the near side and overestimate those to the far side. This would give the impression of a larger cloud distribution in the spiral arm compared to the actual one. 

\subsection{Distance Errors derived from a simulated Milky Way Galaxy}
\label{sec:KD-Actual-Sim}

In this section, we perform the analysis described in the previous one using the actual kinematics derived from the simulation, which includes net radial and azimuthal drifts in the clouds' velocity distribution due to the spiral arm perturbation. We quantify the error introduced by these components in the kinematic distance estimate.

\subsubsection{Section I}

The kinematic distance estimate compared to the actual distance for the clouds in this section is presented in the upper right panel of Figure \ref{fig:kdist-SPAM760}, which indicates that the distance for almost all clumps is overestimated. The effect is better visualised in the distance error ($D_{\mathrm{kin}} - D_{\mathrm{actual}}$) distribution for this section, which is shown in the upper right panel of Figure \ref{fig:SPAM760-dist-error-hist-quads}. The error shows a systematic offset of $\approx 1$ kpc and ranges between $-1$ kpc and $2$ kpc. About $87\%$ of the clouds have distance errors between $0.5$ and $1.5$ kpc. When these values are converted to a fractional error, the range is between $-5$ \% and $16 \%$. The cloud distribution has a net inward radial streaming motion, which contributes more to the line of sight velocity in this section. Because the projection of $v_c$ and $v_R$ on the line of sight points in opposite directions, the projected velocity tends to be lower than the expected from a  circular orbit. This results in an distance overestimates in the kinematic method, which can explain the systematic shift to positive values in the error distribution.

\subsubsection{Section II}

For this section, the upper left panel in Figure \ref{fig:kdist-SPAM760} shows that the kinematic method tends to overestimate the distance and the error increases as the actual distance decreases. The distance error distribution, which is given in the upper left panel of Figure \ref{fig:SPAM760-dist-error-hist-quads}, shows a clear peak near $2$ kpc and a smaller maximum at around $0.7$ kpc and is roughly centred around $1$ kpc. The error ranges between $-1$ and $3$ kpc, with the exception of a few clumps with errors near $-2$ kpc. The fraction of clouds with errors within $0.5$ and $1.5$ kpc is $38\%$, which includes the smaller peak in the distribution. In the range between $1.5$ and $2.5$ kpc, where the larger peak is located, the fraction of clouds is $34\%$. This means that about $72\%$ of the clouds have distance errors in the range between $0.5$ and $2.5$ kpc. Around $23\%$ of the clumps is found in the range between $-0.5$ and $0.5$ kpc. In terms of a fractional error, the distribution ranges from $-58 \%$ up to $80 \%$. This cloud groups lies very near the tangent point, where the contribution of $v_R$ is negligible. However, the clouds are separated in a fast and slow moving group, which can explain the bimodal distribution. The overall shift of the errors to positive values may be explained by the slower net rotation of the cloud distribution as a whole with respect to the circular velocity.

\subsubsection{Section III}

In this section, the distance error has a tendency to grow as the actual distance decreases, as shown in the lower left panel in Figure \ref{fig:kdist-SPAM760}. The error distribution for this section is clearly bimodal with peaks near $1$ and $-0.5$ kpc and has a slight offset from zero. This is presented in the lower left panel of Figure \ref{fig:SPAM760-dist-error-hist-quads}. The fraction of clouds with errors within $-1$ and $0$ kpc is $41\%$, which encloses one of the peaks of the distribution. In the range between $1$ kpc and $2$ kpc, where the second peak is found, the fraction drops to $27\%$. A similar fraction is found in the intermediate range between $0$ and $1$ kpc. In terms of the fractional error, the distribution ranges from $-28 \%$ to $39 \%$. Most of the clouds have errors shifted to negative values. This cloud group is more offset from the tangent point, increasing the contribution of the $v_R$ component in the line of sight velocity. The bimodal distribution can be explained by the two main circular velocity cloud groups.

\subsubsection{Section IV}

\begin{figure*}
	\begin{center}
		\includegraphics[width=0.95\columnwidth]{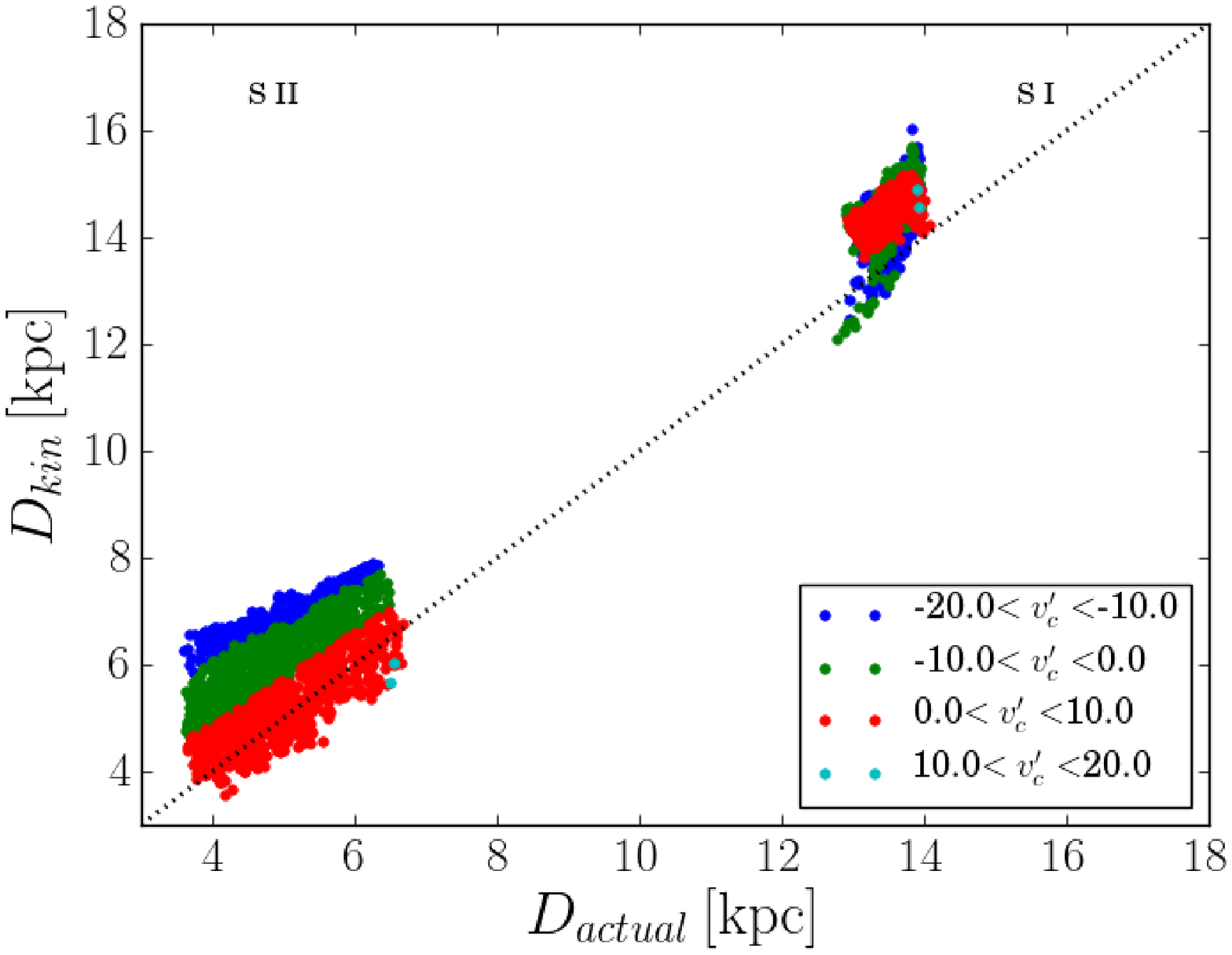}
		\includegraphics[width=0.95\columnwidth]{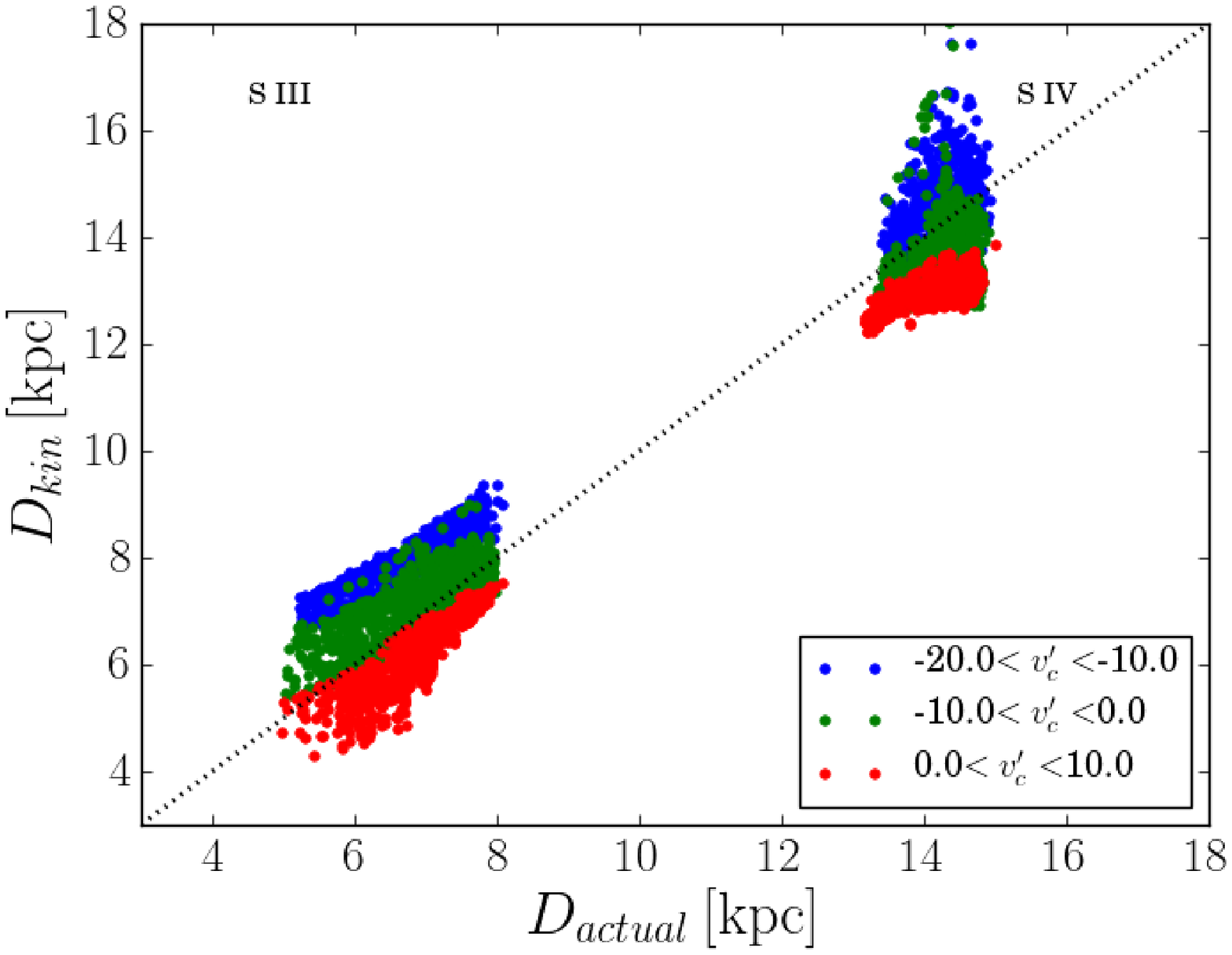}

		\caption{Cloud kinematic distance \dkin~compared to the actual value $D_{\mathrm{actual}}$. The cloud kinematics obtained from the spiral arm simulation, which include both radial and azimuthal streaming motions, where used for these distance estimates. Sections I and II are shown in the \emph{left panel}. Sections III and IV are shown in the \emph{right panel}.}
		\label{fig:kdist-SPAM760}
	\end{center}
\end{figure*}

\begin{figure*}
	\begin{center}
		\includegraphics[width=0.95\columnwidth]{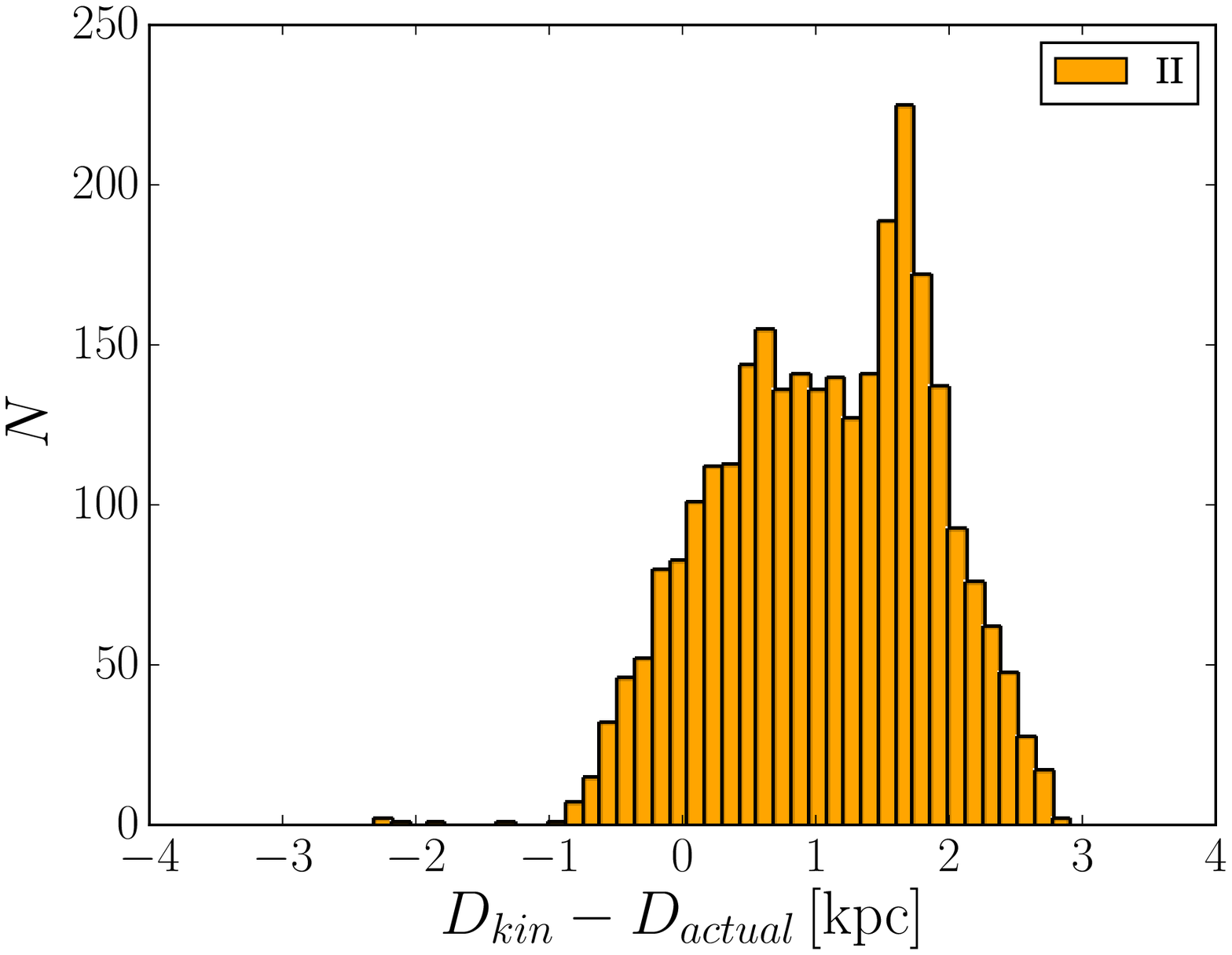}
		\includegraphics[width=0.95\columnwidth]{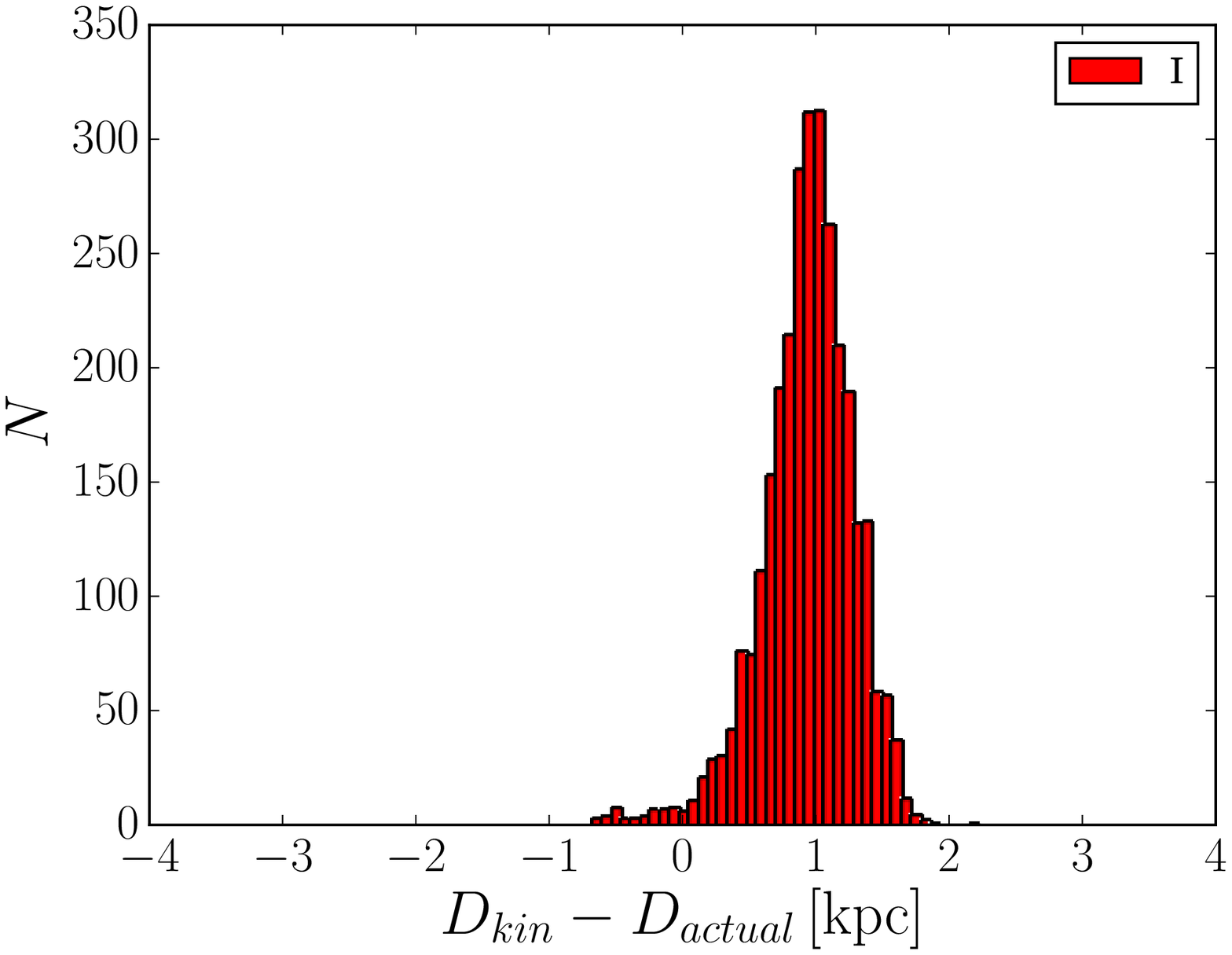}
	
		\includegraphics[width=0.95\columnwidth]{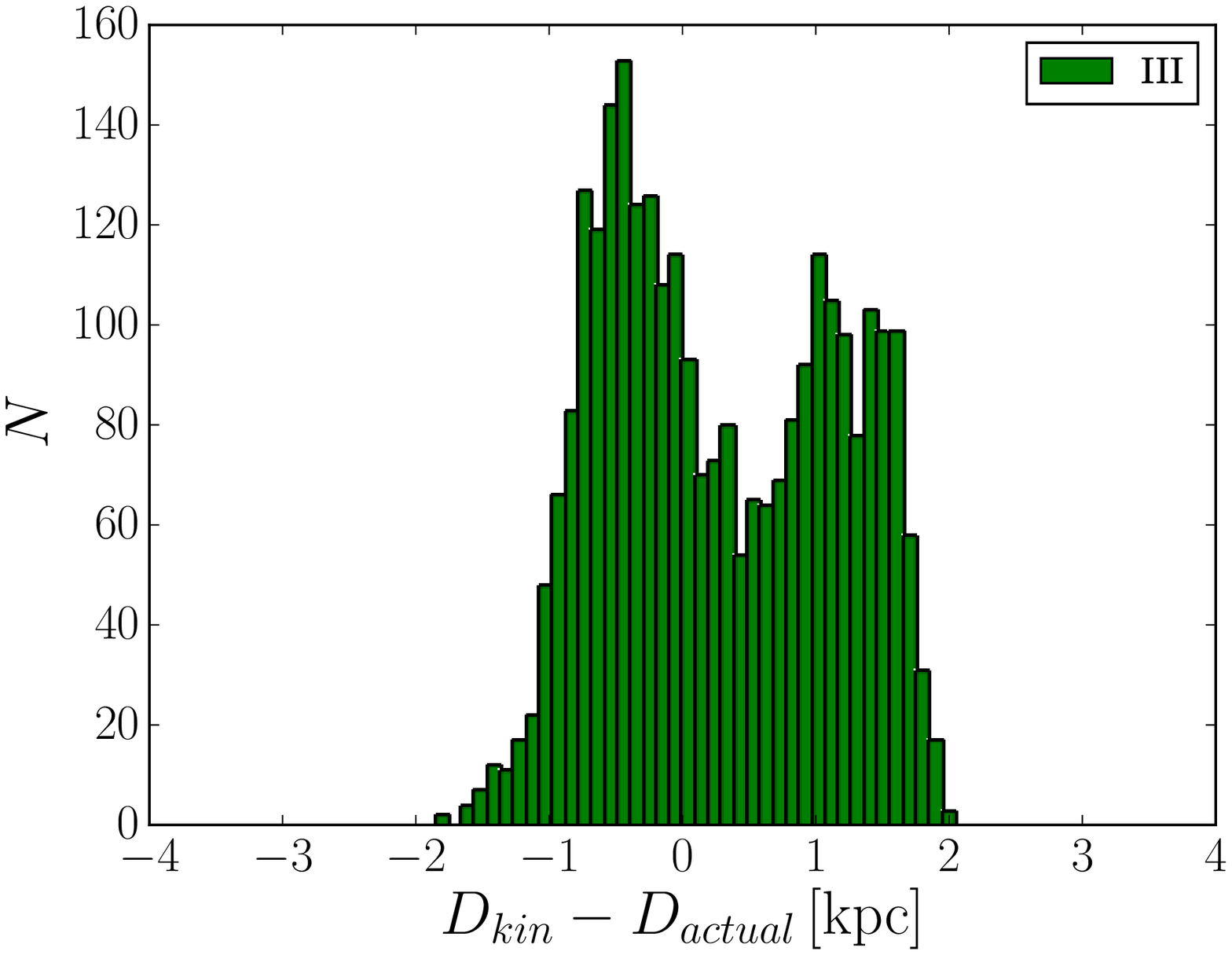}
		\includegraphics[width=0.95\columnwidth]{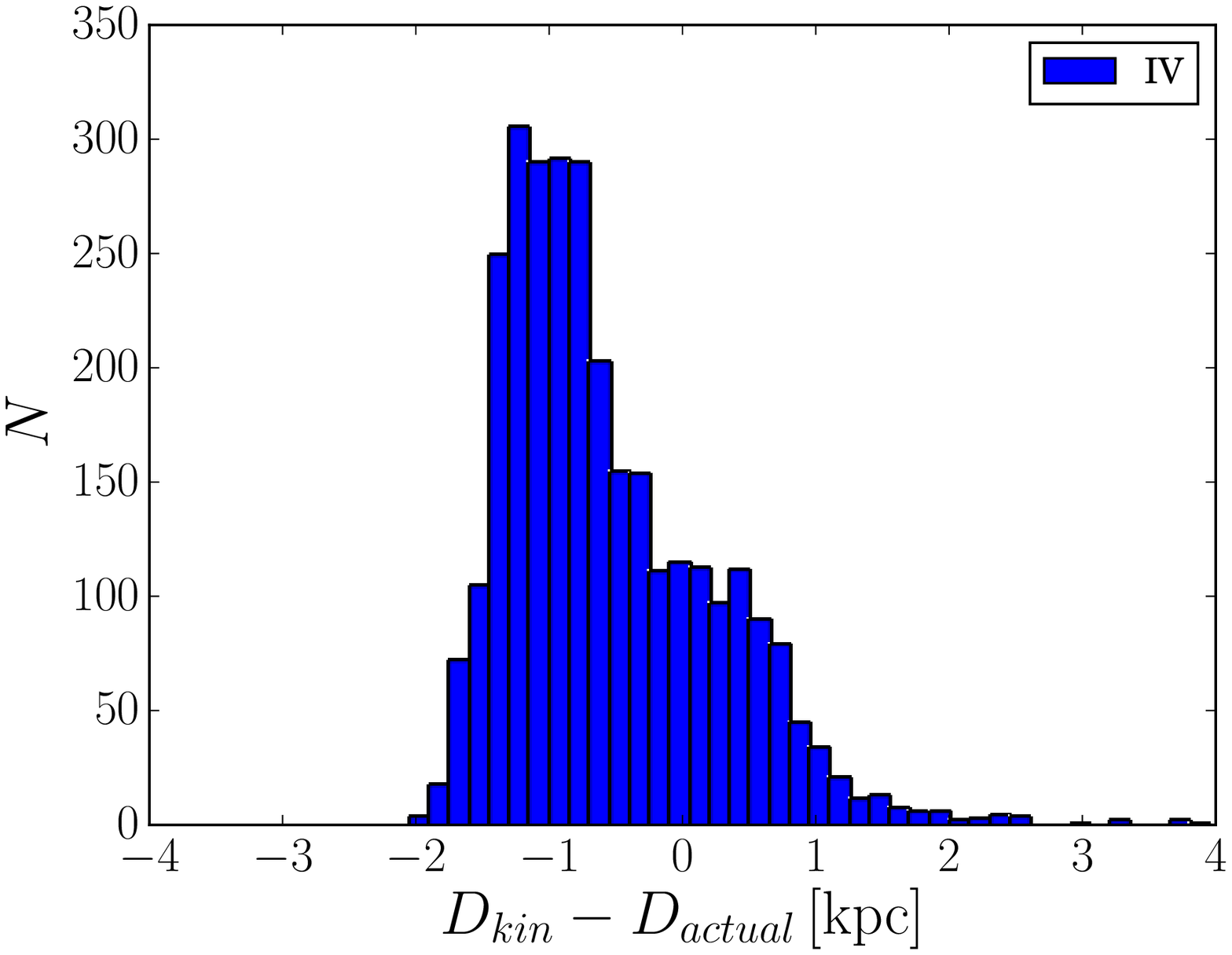}
		\caption{Distance error ($D_{\mathrm{kin}} - D_{\mathrm{actual}}$) distributions divided by section: section I (\emph{upper right}), section II (\emph{upper left}), section III (\emph{lower left}), section IV (\emph{lower right}). These are derived from the actual kinematics in the simulation, which include both radial and azimuthal streaming motions. The net offset in section I and IV is a consequence of a net inward radial motion and the bimodality in section II and II is a result of two main groups in the circular velocity.}
		\label{fig:SPAM760-dist-error-hist-quads}
	\end{center}
\end{figure*}

The distance error grows with increasing cloud distance, as shown in the lower right panel of Figure \ref{fig:kdist-SPAM760}. In the error distribution, it has a systematic offset of $\approx -1$ kpc (see Figure \ref{fig:SPAM760-dist-error-hist-quads}). The distance error ranges from $-2$ kpc up to $4$ kpc for a few clouds. We find that $58\%$ of the clumps have errors within $-1.5$ and $-0.5$ kpc and $34\%$ are in the range between $-0.5$ and $1$ kpc. In terms of the fractional error, the distribution ranges from $-13 \%$ to $27 \%$. In this case, the projection of $v_\phi$ and $v_R < 0$ point in the same direction along the line of sight. This tends to make the line of sight velocity more negative compared to the value expected from the circular velocity. The result is a tendency to underestimate the distance with the kinematic method, which explains the shift of errors to negative values.

\subsubsection{Recovered Positions}

\begin{figure*}
	\includegraphics[width=0.95\columnwidth]{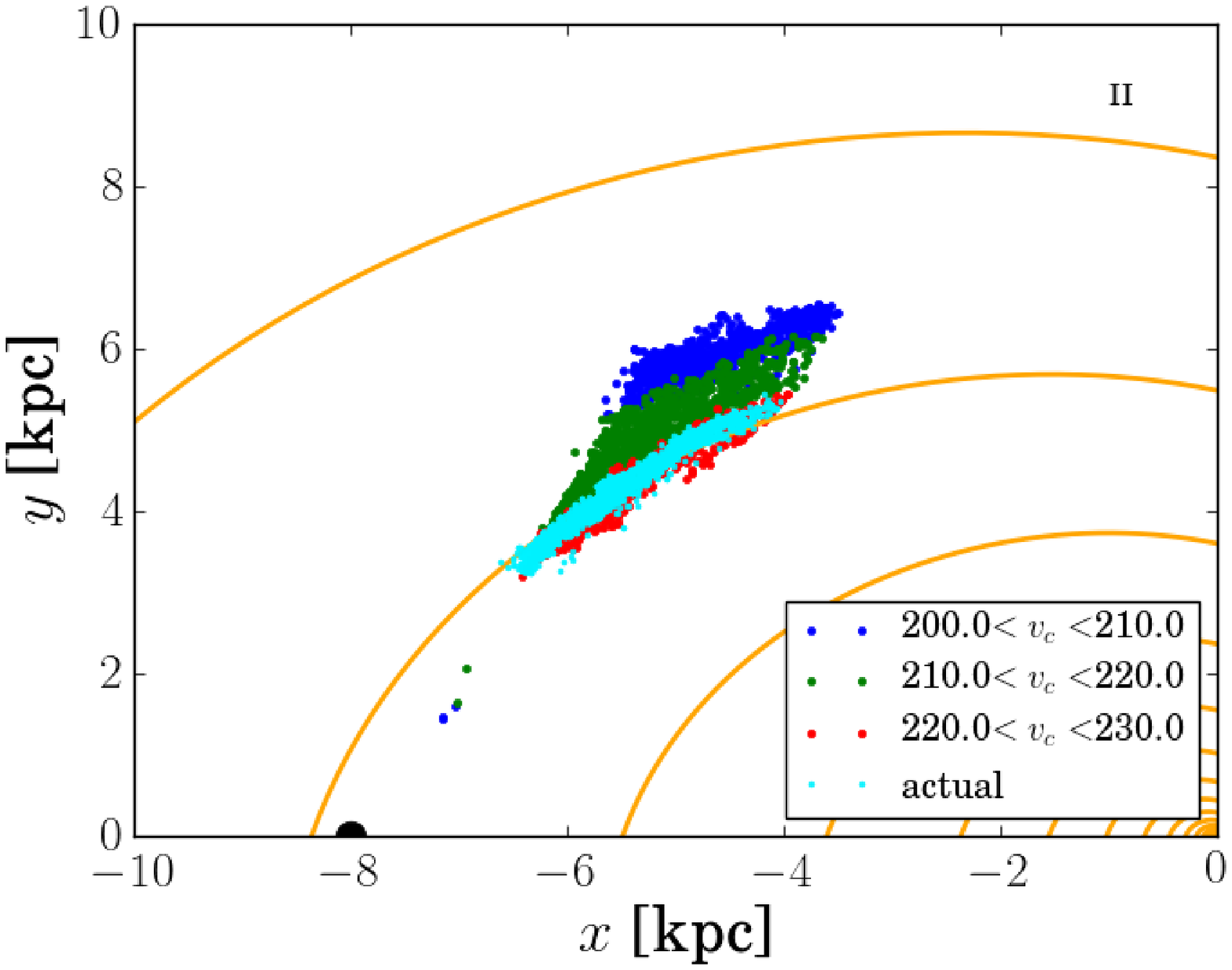}
	\includegraphics[width=0.95\columnwidth]{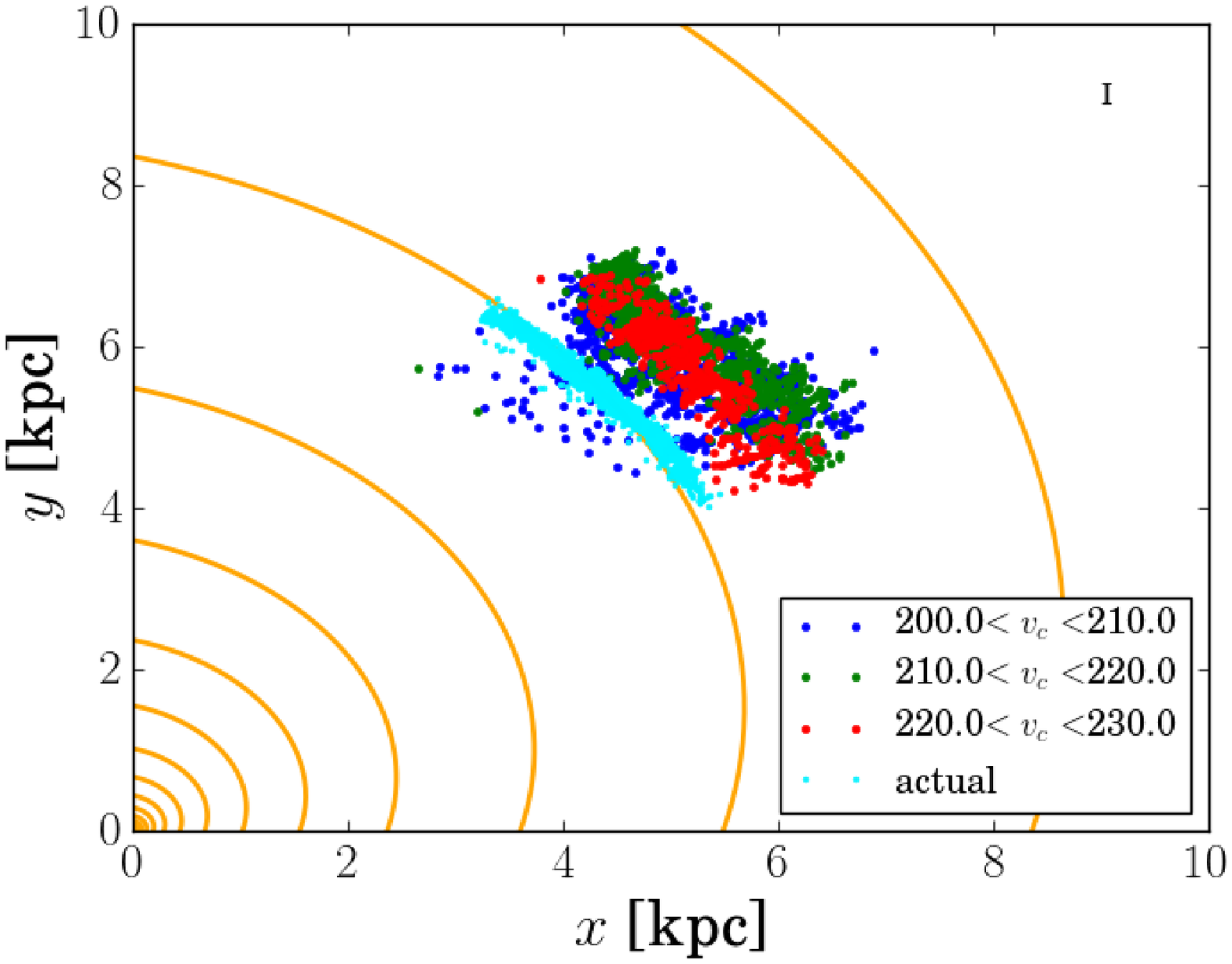}

	\includegraphics[width=0.95\columnwidth]{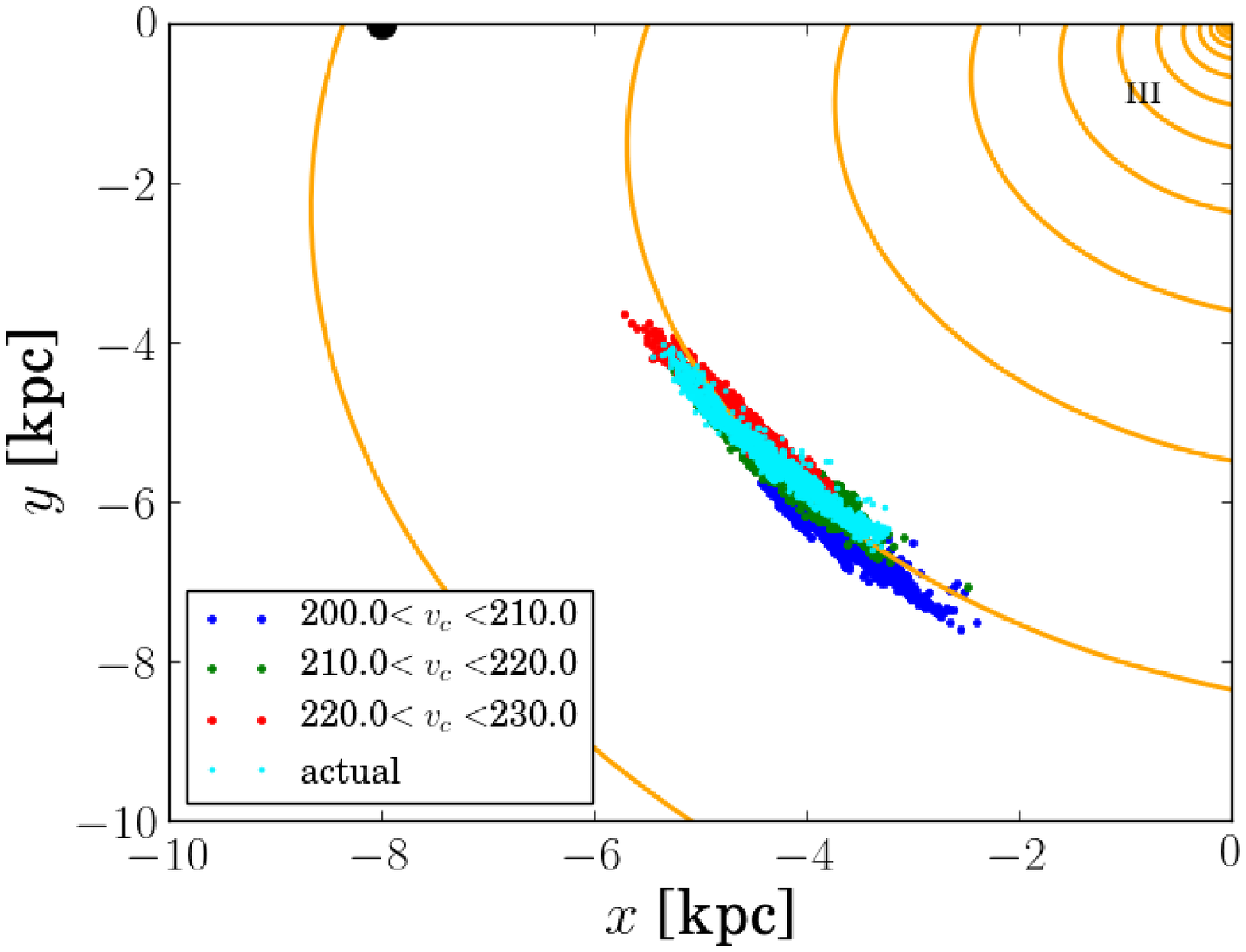}
	\includegraphics[width=0.95\columnwidth]{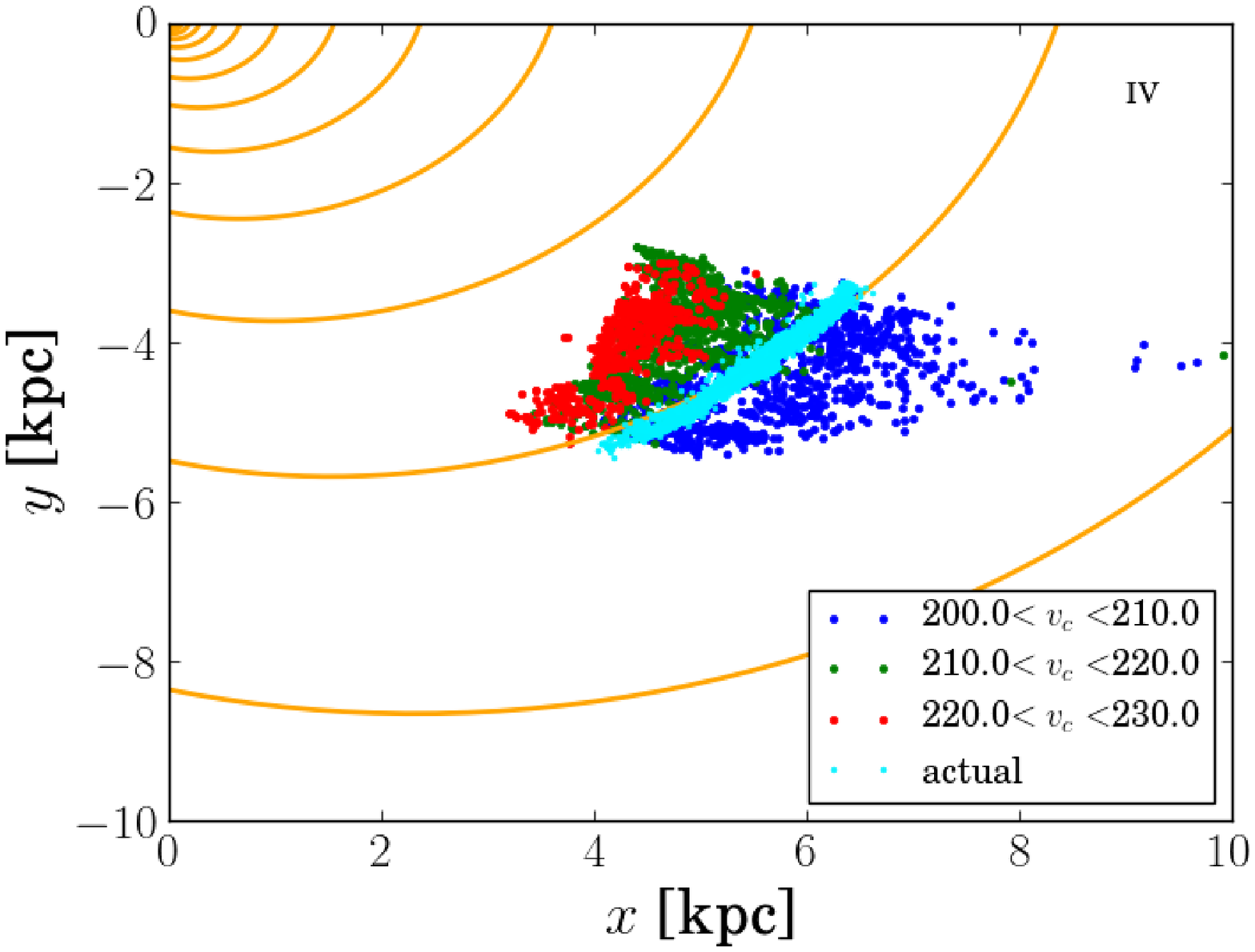}
	\caption{Recovered position maps colour coded by the cloud's azimuthal velocity component obtained in the simulation. The solid curves represent the spiral arms tracing the actual cloud distribution. These results show the effect of the error introduced in the kinematic distance estimate by net streaming components due to the spiral arm.}
	\label{fig:recoveredpos-SPAM760}
\end{figure*}

Figure \ref{fig:recoveredpos-SPAM760} shows the positions recovered by an imaginary observer given the obtained kinematic distances compared to the spiral arms tracing the original distribution. 

In section I, the recovered cloud distribution is shifted to a farther distance than the original arm. In section II, the slower clouds have an overestimated distance, resulting in an apparent net shift and increase in spatial dimensions of the distribution with respect of the original arm. The cloud distribution in section III is more elongated compared to the original. Finally in section IV, it is possible to see a significant distortion of the cloud positions with a few of them having recovered positions closer to a different spiral arm.

Compared to the results of \S \ref{sec:effect-vdisp-kd}, these error distributions show that net streaming motions in the cloud distribution introduce systematic errors as large as $1$ kpc in the distances derived using the kinematic method. The cloud-to-cloud velocity dispersion also propagates into a distribution of distance errors.

\section{Discussion}
\label{sec:discussion}

Our simulations show significant streaming motions through the spiral arms. Although the net motion is from the trailing to the leading edge, the peculiar velocities, compared to the local circular velocity, appear to show motion in the opposite direction. The typical peculiar motions we find are $\approx 15$ \kms, which are comparable to those observed in \citet{Reidetal2014}

Recent surveys by \citet{Choietal2014}, \citet{Hachisukaetal2015} and \citet{SatoWuImmeretal2014}, find that star forming regions have a net motion toward the Galactic centre in the Outer, Perseus and Scutum Arms, respectively. This is consistent with our simulation. \citet{Hachisukaetal2015} and \citet{SatoWuImmeretal2014} find average galactocentric motions of $10$ \kms~and $8$ \kms. The average $v_R$ of clouds in our simulation is close to this value. However, the analyses of \citet{XuLiReidetal2013} and \citet{WuSatoReidetal2014} of the Local and Sagittarius Arms, respectively, show that the average galactocentric motion is close to zero, suggesting some arm to arm variations. We note that the derivation of peculiar motions is sensitive to the Solar Motion used (e. g. \citealt{McMillanBinney2010}).

In the circular component, \citet{Reidetal2009} find that star forming regions are rotating about $15$ \kms~more slowly than the rotation curve, based on the Solar Motion of \citet{DehnenandBinney1998}. However, such measurement is highly sensitive to this reference (e.~g. \citealt{McMillanBinney2010,Honmaetal2012}). An updated analysis in \citet{Reid2012} based on the Solar Motion of \citet{SchonrichBinneyDehnen2010} shows a lower lag of $\approx 6$ \kms. Recent works suggest smaller values: $\approx 5$ \kms~for the Local Arm \citep{XuLiReidetal2013}, $\approx 4$ \kms~for the Scutum Arm \citep{SatoWuImmeretal2014} based on the Solar Motion of \citet{SchonrichBinneyDehnen2010} and \cite{Reidetal2014}, respectively. However, \citet{WuSatoReidetal2014} find an average motion $\approx 3 $\kms~faster than the rotation curve in the Sagittarius arm. These values are in agreement with the average kinematics in our simulation. However, we find a group of clouds lagging $\approx 15$ \kms~behind the rotation curve, corresponding to clouds leaving the arm.

Streaming motions may be sensitive to the assumed spiral arm potential. $N$-body simulations by \citet{Babaetal2009}, where arms have a more transient evolution and varying morphology, show that star-forming regions are not necessarily moving slower than the rotation curve and peculiar motions tend to be random, which is different to our results. However, they find motions as large as $20 \%$ of the circular velocity. This agrees with our simulation, but we find a slower typical motion. Using a gas orbit analysis approach, \citet{Honmaetal2015} find that gas in a spiral arm potential can move slower and faster than the rotation curve depending on the galactocentric radius, which is comparable to our simulation.

In terms of kinematic distances, \citet{Reidetal2009} found that the difference between the kinematic estimate and the parallax distance can be larger than a kpc, which is consistent with the range of errors found in our analysis. \citet{Reidetal2009} propose a version of the kinematic distance method which includes the systematic motions derived from their observations. It showed some improvement, but for a few sources the error was still large \citep{Reidetal2009}.

\citet{Roman-Duvaletal2009} estimated the distance uncertainty by adding perturbations to the line of sight velocity in the range between $-15$ and $15$ \kms at galactic longitudes $l = 20^\circ$ and $l = 40^\circ$ and find errors lower than $30\%$. In our simulations, the clouds in sections I and IV have galactic longitudes $l \approx \pm 20^\circ$ and have relative errors not larger than $30\%$, which is consistent with the analysis of \citet{Roman-Duvaletal2009}. However, our results suggest that when using the kinematic distance estimate, it may not be safe to assume that the cloud's circular velocities are distributed symmetrically around the rotation curve.

Both \citet{Andersonetal2012} and \citet{Wienenetal2015} calculate the distance uncertainty by taking into account a line of sight component of $\pm (7 - 8)$ \kms. \citet{Andersonetal2012} find a relative uncertainty less than $20\%$. These values may be reasonable to estimate the error introduced by a cloud-to-cloud dispersion, but is not taking into account the systematic error that may be introduced by streaming motions in the cloud distribution.

From the point of view of simulations, \citet{Gomez2006} studied the errors in the kinematic distance in a global model representative of the Milky Way and concluded that the distance error increases in lines of sight in the direction of the spiral arms and can be larger than a kpc, clearly affecting the geometry of the arm reconstructed from the derived distances. \citet{Babaetal2009} also reach to a similar conclusion. The distance errors we find are comparable to those reported in \citet{Gomez2006} and \citet{Babaetal2009}, and our recovered position maps also verify the fact that the inferred spiral structure is considerably distorted. Our simulation has the resolution to follow in more detail the kinematics of individual gas clouds in the vicinity of the spiral arm which allows us to quantify the errors introduced by more localised streaming motions. However, we are limited in the sense that we have simulated a particular region and assumed that these kinematics equally apply to the other arms. In a more realistic Galaxy model, there may be some arm to arm differences that can introduce additional variations.

\section{Conclusions}
\label{sec:conclusions}

We identify dense gas clouds in a high-resolution hydrodynamical simulation of a region of gas flowing through a spiral arm and we study the motions of these clumps with respect to the galaxy. We use the kinematics of these clouds to analyse the errors that these motions can introduce in the kinematic distance method.

The spiral arm perturbation introduces net radial and azimuthal streaming motions in the gas motions. Although the net motion is from the trailing to the leading edge, this results in peculiar motions toward the galactic centre and opposite to the local circular velocity. This also introduces systematic errors of $\approx 1$ kpc in the kinematic distance. The cloud-to-cloud velocity dispersion introduces an additional scatter in the error. 

For an observer attempting to map the clouds' positions based on the derived distances, these effects result in distorted structures with systematic offsets with respect to the actual structure. In our simulations, most of the distance errors are in the range of $\pm 2$ kpc. For surveys using the kinematic distance estimate, this increases the likelihood of misdiagnosing spiral arms affecting our understanding of where molecular clouds and star formation regions are with respect to spiral structure.

\section*{Acknowledgements}

We thank the referee, Mark Reid, for valuable suggestions that improved this paper.
FGRF and IAB gratefully acknowledge support from the ERC ECOGAL project,
grant agreement 291227, funded by the European Research Council under ERC-2011-ADG.
This research has used NASA's Astrophysics Data System. This work used computational
resources of the DiRAC Complexity system, operated by the University of Leicester IT
Services, which forms part of the STFC DiRAC HPC facility. We used the code SPLASH
by \citep{Price2007} for visualisation of SPH simulation data. We are also grateful to
Claudia Cyganowski, Maya Petkova, Duncan Forgan, William Lucas for their helpful discussions 
on this paper.




\bibliographystyle{mnras}
\bibliography{references_lib} 








\bsp	
\label{lastpage}
\end{document}